\documentclass[aps,prb,twocolumn,floatfix]{revtex4-2}
\usepackage{amsmath}
\usepackage{amssymb}
\usepackage{graphicx}
\usepackage{amsfonts}
\usepackage{physics}
\usepackage{comment}
\usepackage{float}
\usepackage{natbib}
\usepackage{color}

\begin{document}
\title{\Large{Resonant tunneling diodes in semiconductor microcavities: \\ modeling polaritonic features in the THz displacement current}}
\author{Carlos F. Destefani$^1$}
\altaffiliation[email: ]{carlos.destefani@uab.es}
\author{Matteo Villani$^1$}
\author{Xavier Cartoix\`{a}$^1$}
\author{Michael Feiginov$^2$}
\author{Xavier Oriols$^1$}
\altaffiliation[email: ]{xavier.oriols@uab.es}
\affiliation{$^1$ Department of Electronic Engineering, Universitat Aut\`onoma de Barcelona, 08193 Bellaterra, Barcelona, Spain\\
$^2$ Department of Electrical Engineering and Information Technology, Technische Universität Wien, 1040 Wien, Austria}
	
\begin{abstract}
We develop in this work a simple qualitative quantum electron transport model, in the strong light-matter coupling regime under dipole approximation, able to capture polaritonic signatures in the time-dependent electrical current. The effect of the quantized electromagnetic field in the displacement current of a resonant tunneling diode inside an optical cavity is analyzed. The original peaks of the bare electron transmission coefficient split into two new peaks due to the resonant electron-photon interaction, leading to coherent Rabi oscillations among the polaritonic states that are developed in the system in the strong coupling regime. This mimics known effects predicted by a Jaynes-Cummings model in closed systems, and shows how a full quantum treatment of electrons and electromagnetic fields may open interesting paths for engineering new THz electron devices. The computational burden involved in the multi-time measurements of THz currents is tackled by invoking a Bohmian description of the light-matter interaction. We also show that the traditional static transmission coefficient used to characterize DC quantum electron devices has to be substituted by a new displacement current coefficient in high-frequency AC scenarios.   
\end{abstract}

\maketitle

\section{Introduction}
\label{intro}

In the transition from micro- to nano-electronics a revolution appeared thirty years ago in the way electron devices were modeled. The classical treatment of electrons had to be substituted by a quantum view to account for new phenomena like electron tunneling and energy quantization. Such a transition was envisioned by Landauer as early as 1957 \cite{Landauer_1957}. Nowadays, as electron devices reach low nanometer scales, most typical electron device models have adopted a quantum treatment for electrons \cite{BIT,PhonTS,Dalle_2020,Tiber_CAD,nextnano,Klimeck2,Rossi1,Rossi2,Ferry_2018,Jonasson_2015,Querlioz_2013,Klimeck3,Klimeck4, Green,Green2}, but still only Coulomb law is usually considered. Such models couple the Gauss law into the quantum electron transport equation within an electrostatic approximation. For nanoscale devices and frequencies in the THz gap ($\approx 0.1$-$30$ THz)  \cite{BIT,PhonTS,Dalle_2020,rev_manip_THz,Tataria,Preu,Knezevic1} such electrostatic approximation is no longer valid. Such a frequency gap has been explored by a growing number of technologies, from frequency down-conversion of photon sources \cite{Nagatsuma_2016_1,Hovenier,Gunn_diode,Koenig_2013,Nagatsuma_2016_2,Chinni_2018,Jia_2017} to frequency up-conversion of electron sources \cite{Nagatsuma_2018_2,Nishizawa_2008,Acharyya_2012,Kalid_2014,lewis_2014,Brown,Asada_2021,Asada1,Asada_2005,Wasige_2015,Arzi_Prost_2015} in applications from medicine \cite{Yeo_2011,Brown_2018,Taylor_2014,Nishizawa_2005,Yeo_2019} to telecommunication \cite{Song_2021,Koenig_2013,Nagatsuma_2016_2,Jia_2017,RTDOrev}. It has also been experimentally demonstrated that, at high-frequency regime, displacement currents play dominant role in Resonant Tunneling Diodes (RTDs) \cite{Fei07,Fei00,Fei00a,Fei01,Fei11b} and in RTD oscillators \cite{Fei11a,Fei12}. 

It is nowadays evident that the study of THz electron devices requires, not only the time-dependent Gauss law, but also  a full treatment of Maxwell equations. For example, the role of external magnetic fields on dynamics of electron devices can straightforwardly be incorporated and it is well studied \cite{Jensen2,Dollfus3,laura22}. In traditional electron devices, the study of the coupling between electron transport models and classical electromagnetic fields has generated plasmonic devices, in which fluctuations of charge density induce oscillations of the electromagnetic fields and vice-versa \cite{Dyakonov,Lu,Deng}, that have been considered as electron sources and detectors at the THz gap \cite{Bhardwaj}. 

At THz the photon energies may become comparable to the electron energies, so that one can wonder whether the quantization of the electromagnetic fields, in addition to the electrons, is also required. When electrons and photons can be treated as independent entities, the quantization of the electromagnetic fields has been successfully included in many quantum simulators \cite{Rossi1,Rossi2,Ferry_2018,Jonasson_2015,Querlioz_2013,Klimeck3,Klimeck4, Green,Green2}, where the effects of the fields are translated into collisions in the electron dynamics, with electrons absorbing or emiting a photon according to Fermi Golden Rule. However, in scenarios where a strong light-matter coupling is induced, such a perturbative approach can no longer be applied, and a quantization of both electrons and fields is required. The coupling strength among light and matter is usually given by the factor $\zeta = \omega_r / \omega$, in which the Rabi frequency $\omega_r$ embodies such coupling, and the field frequency $\omega$ may or not be in resonance with some typical electron frequency of the device $\omega_e=(E_1-E_0)/\hbar$, with $E_{0,1}=\hbar \omega_{0,1}$ two given electron levels. At very weak coupling ($\zeta \ll 0.1$) perturbative techniques mentioned above are enough. At weak coupling ($\zeta \le 0.1$) approaches derived from the Rabi models family, which typically start by dealing with a two level system for the matter and a single mode for the field, are able to capture the light-matter entanglement and have vastly been used in different platforms. At strong coupling ($\zeta \le 0.5$), such models may still qualitatively handle experimental features, although quantitatively they find limitations due to their simplifications. Recent experimental achievements \cite{ultrastrong_reviewNAT,ultrastrong_reviewRMP} have reached the ultra strong ($\zeta \le 1$) and deep ultrastrong ($\zeta \ge 1$) regimes, claiming for new theoretical models (such $\zeta$-values are estimations).

Theoretical models for light-matter interaction overall start from a minimal-coupling Hamiltonian with respect to the vector potential, and proceed either in Coulomb gauge, by employing a transformation that rewrites the field operators as to include the diamagnetic shift, or in Power-Zienau-Woolley (PZW) gauge \cite{PZW_deriv, PZW_2D}, where the Hamiltonian is given in terms of the fields themselves instead of the vector potential, and that also takes care of the diagmagnetic term and introduces the dipole self-energy. There has been some controversy at a fundamental level, where it was claimed \cite{PZW_probs,PZW_probs_reply} that gauge invariance is not obeyed among these gauges, which was lately refuted \cite{PZW_err,PZW_err2}. However, it is true that, in practical calculations, when one needs to restrict the subsets of electron and photon states, such invariance is broken, most importantly at strong couplings which could require larger basis sets. Methods have been proposed to handle such a problem \cite{gaugebreakdown}, so that gauge invariance may exist even within restricted subspaces as typical in the Rabi models family \cite{PZW_gauge,PZW_ambigs}, or in an external field \cite{PZWno_timedep}. They also claim that, at strong couplings, the Coulomb gauge is more affected by the basis size, while others claim that a truncated basis set \cite{gaugestrong,gaugestrong2}, even in a Jaynes-Cummings model \cite{cavity_A2,gauge_ultrastrong_JC}, could still be used under strong couplings by employing a specific unitary transformation. 

The correlated light-matter states are the so-called polaritons, which come in different `flavours' depending on the matter entity (excitons, electrons, plasmons). Polaritons \cite{polarit_dev,singlep_hydro} have been widely studied in several platforms: vacuum-field enginneering of materials \cite{QWsplit92,vacuum_chem1,vacuum_chem2,vacuum_chem3,vacuum_chem4}, strongly coupled quantum dot emitters in vacuum field \cite{QDvacuum,QDvacuum2}, control of chemical reactions \cite{polariton1,polariton2}, light-induced states of matter \cite{induced2,gaugestrong,induced4}, solid state logic gates based on external pump of excitonic polaritons \cite{gatespolariton,gatespolariton2,gatespolariton3,gatespolariton4}; the latter flavour allowed the realization of Bose-Einstein condensates in semiconductor \cite{BEC_exc_pol,BEC_exc_pol2,BEC_exc_polR} and organic \cite{BEC_org} cavities in the strong coupling regime. Other than exciton polaritons, which carry information from both conduction and valence bands, there also exists intersubband polaritons in vacuum or external fields \cite{ISBpol1,ISBpol2,ISBpola,ISBpolb,ISBpol3,ISBpoldip,ISBpol4}, in which only conduction band electrons play a role. Most of these works operate at mid-infrared/THz regime \cite{THZ_QCL,THZ_review,ISB_THz,ISB_ultrastrong_THz,ultrastrong_time}. The overall goal is, since light is not akin to interactions, to correlate it with matter as strongly as possible so one can manipulate matter more easily. 

In a different front, selfconsistent ab-initio procedures for many-particle systems at strong light-matter coupling have been worked, for example, through quantum electrodynamical density functional theory (QEDFT) \cite{QEDFT_bridge_2014,QEDFT_refereeexp_2018,QEDFT_noGS_2018,QEDFT_selfmaxwell_2019,QEDFT_refereePZW_2020,pnas,QEDFTno_refereeNOcurrent_2022,TDDFT_2022}, which also tries to probe DC dynamics \cite{QEDFT_selfmaxwell_2019,QEDFTno_refereeNOcurrent_2022} in a selfconsistent manner through the linear Kubo approximation. On the other hand, modifications of transport properties as induced by a strong light-matter coupling have been addressed in 2DEGs \cite{conductivity2DEG}, quantum hall systems \cite{conductivity2DEGnon,conductivitylandau}, and organic semiconductors \cite{conductivityorgan}. Developing transport models at strong field couplings \cite{QCDtunnel,QCDtunnel0}, where electron dynamics is so affected, for example, by the quantized electromagnetic field of a cavity or by a high-frequency external signal, remains an open challenge for the rise of the next generation of THz electron devices. 

\begin{figure}
\includegraphics[scale=0.6]{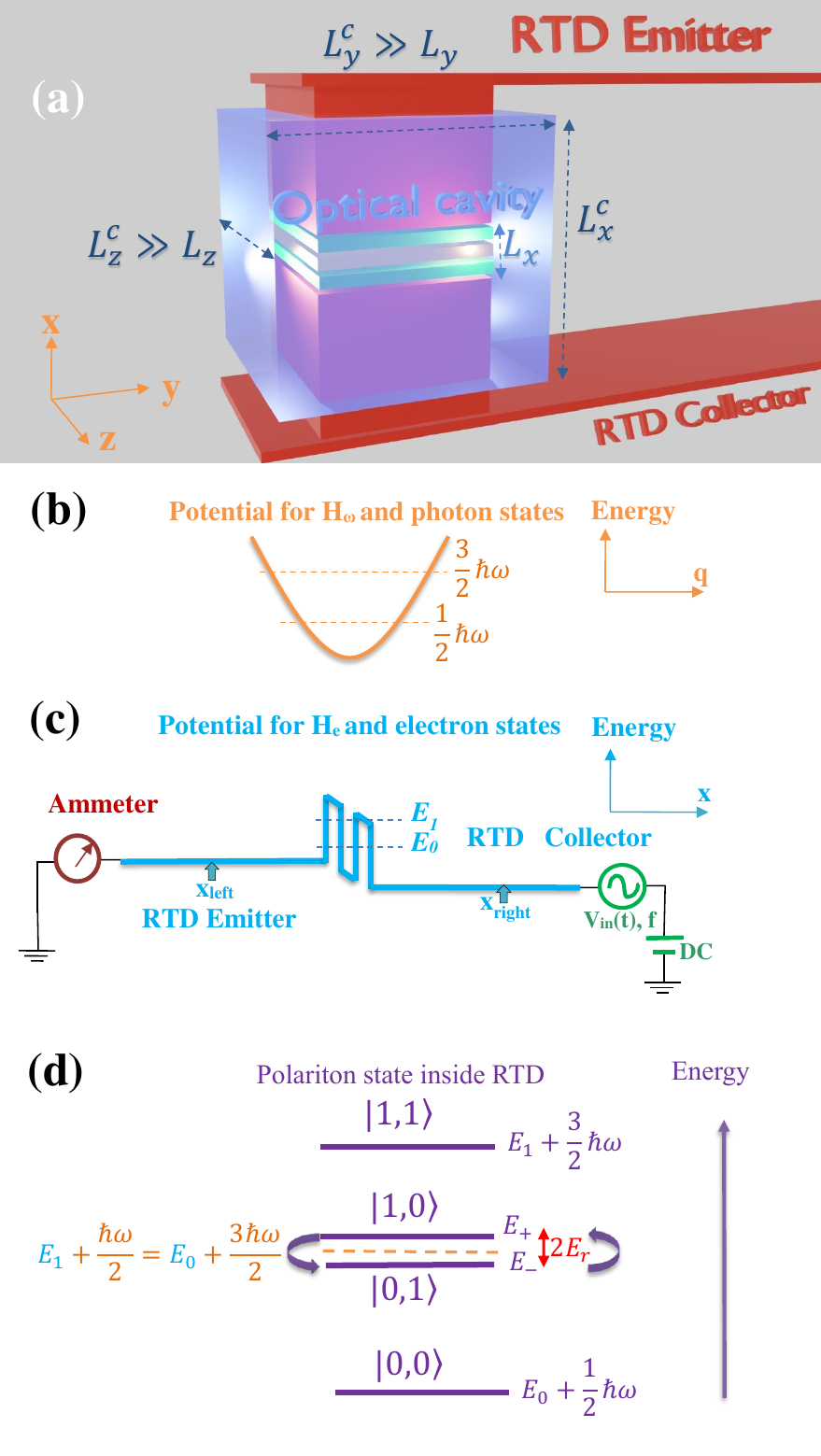}
\caption{(a) $3$D spatial representation of the transport through a RTD whose active region is inside a cavity, and whose transport direction size is much smaller than the lateral sizes, $L_y,L_z \gg L_x = 16$ nm. (b) zero-photon $|0 \rangle$ and single-photon $|1 \rangle$ states for the quantized single mode cavity field with energies $\hbar\omega/2$ and $3 \hbar\omega/2$. (c) $1$D-view of the RTD device showing ground $|0 \rangle$ and first excited $|1 \rangle$ electron states with energies $E_0$ and $E_1$, so that $\omega_e=(E_1-E_0)/\hbar$ is in the THz range; the light-matter interaction is effective only inside the active region, while a much larger simulation box is used to deal with open boundary conditions, with $x_{left,right}$ indicating the positions where wavepackets are initialized. (d) $|\text{electron},\text{photon} \rangle$ states inside the RTD/cavity in the resonant strong coupling regime: state $|0,0 \rangle$ almost unnafected; polaritonic states formed out of $(|0,1 \rangle \pm |1,0 \rangle)/\sqrt{2}$ split by $2 E_r = 2 \hbar \omega_r$ in comparison to the degenerate decoupled energies (dashed line); state $|1,1 \rangle$ would create another polariton subspace, in a larger basis set, with state $|0,2 \rangle$. An external battery yielding a signal $V_{in}(t)$ with frequency $f$, as well as an ammeter, are also shown in (c). The dimensions of the cavity $L_{x,y,z}^c$ are tens of $\mu$m as to yield a frequency $\omega \approx \omega_e$.}
\label{volume}
\end{figure}

The main goal of our work is to develop a simple model for THz electron devices in the strong coupling regime, able to qualitatively capture polaritonic signatures in transport features. We consider a typical RTD device where only electrons in the conduction band become relevant for transport, with lateral dimensions $L_{y,z}$ much larger than the transport dimension $L_x=16$ nm (Fig. \ref{volume}(a)). As such, the three electronic coordinates are treated independently and transport is described by a $1$D effective mass equation with a double barrier potential profile, yielding the electron levels as depicted in Fig. \eqref{volume}(c), whose splitting $\hbar \omega_e$ is in the THz range \cite{RTD_nano,Asada_2021,Asada1,Asada_2005}. Electrons are defined as time-dependent (conditional) wavepackets that are injected from emitter to collector and, along the way, cross a semiconductor microcavity \cite{RTD_cavityexp,RTD_polaritonexp,RTD_polaritonPZW} having a quantized single mode $\omega=\omega_e$ resonant electromagnetic field (Fig. \ref{volume}(b)). The sizes of the cavity $L_{x,y,z}^c$ are much larger than the respective RTD sizes, so that the dipole approximation can be used for the vector potential; we take $L_{x,y,z}^c$ on the order of tens of $\mu$m as to yield a frequency mode in the THz regime, such that indeed $\omega \approx \omega_e$. We consider that the coherent light-matter interaction occurs only inside the RTD active region because outside, in the bulk, the photon energy ($\approx $ meV) is too small compared to the bandgap energy ($\approx $ eV) to generate interband transitions (with conservation of energy), and the photon momentum is too small compared to the electron momentum to generate intraband collisions (with conservation of momentum). In this picture electrons and photons evolve independently in the bulk, and interact only in the active region. The $|\text{electron},\text{photon}\rangle$ states in Fig. \eqref{volume}(d) anticipate that, without coupling (dashed line), states $|0,1\rangle$ and $|1,0\rangle$ are degenerate, while the coupling creates the polariton subspace by inducing an avoided crossing among such states, with splitting $E_+-E_-=2 E_r =2 \hbar \omega_r$. We first calculate the static RTD transmission coefficient $T$. In a second stage, we apply an external AC signal to the RTD profile and compute the total (particle plus displacement \cite{Arzi_Prost_2015,Feiginov_2001,D_mine,D_conference,Feiginov_2001}) current by defining a new displacement current coefficient $D$; we anticipate that $D$ substitutes $T$ at operating frequencies higher than the inverse of the electron dwell time in the active region. Our simplified model captures the polaritonic nature of strong resonant couplings in the transport features of such devices.

The structure of the paper is: Sec. \ref{ham} discusses the Hamiltonians, first stating the Coulomb gauge and then deriving the PZW/Dipole gauge, with a semiclassical model also given; Sec. \ref{dis} employs the Ramo-Shockley-Pellegrini theorem to derive both transmission coefficient $T$ and displacement current coefficient $D$ by interpreting the many-body wave function as a flux of Bohmian conditional wave functions; Sec. \ref{WB} shows the evolution equations under needed approximations for a simplified initial qualitative model; results for static $T$ and dynamic $D$ coefficients are in Sec. \ref{res}; conclusions are in Sec. \ref{con}.

\section{Hamiltonians for light-matter interaction}
\label{ham}

We summarize in this section the two main Hamiltonians \cite{cohen1997photon,grynberg,scully} used to handle light-matter interaction in a many-electron system interacting with a many-mode quantized electromagnetic field, under the dipole/long wavelength approximation. First we obtain the Coulomb gauge Hamiltonian $H^{(C)}$, and then apply a unitary transformation that embodies the diamagnetic term as to obtain a new Hamiltonian $\tilde{H}^{(C)}$. Then we apply the PZW gauge transformation, which takes care of the diamagnetic term and introduces the dipole self-energy, to get the Dipole Hamiltonian $H^{(D)}$; we use such a name for $H^{(D)}$ because we apply the PZW transformation \textit{after} the long wavelength approximation, while the original full PZW transformation is applied \textit{before}, and so it embodies extra magnetic and higher order terms that we neglect in $H^{(D)}$ \cite{QEDFT_refereePZW_2020}. A semiclassical approximation is also given at the end. The Dipole Hamiltonian $H^{(D)}$ is used as the starting point to the Bohmian conditional Hamiltonian $H^{(D)}_{xq}$ in Sec. \ref{Hbohm}, which is an effective way of solving the many-electron many-mode system in terms of equations of motion for a single-electron single-mode problem. Approximations needed in $H^{(D)}_{xq}$ as to handle practical implementations are discussed in Sec. \ref{WB}.

\subsection{Coulomb gauge Hamiltonian}
\label{Hcoulomb}

The minimal-coupling Hamiltonian in the Coulomb gauge, for $N$ confined electrons interacting with a quantized electromagnetic field, in the non-relativistic limit and excluding spin and magnetic interactions, is given by $H = H_0 + H_{\omega 0}$, with \cite{QEDFT_noGS_2018,cohen1997photon}
\begin{equation}
H_0 =  \sum_{j=1}^N \left[ \frac{1}{2m_e}\left( i \hbar\mathbf{\nabla}_j + e \mathbf{A} (\mathbf{r}_j) \right)^2 + eU(\mathbf{r}_j) + V(\mathbf{r}_j) \right],
\label{minimalH}
\end{equation}  
where the scalar potential $U(\mathbf{r}_j)=e(8\pi\epsilon_0)^{-1}\sum_{k=1}^N (|\mathbf{r}_j-\mathbf{r}_k|)^{-1}$ describes the electron-electron interaction ($\epsilon_0$ is the vacuum permittivity), $V(\mathbf{r}_j)$ is some external potential, $m_e$ the (effective) mass of the electrons, and $e=-|e|$ the electron charge, while the field Hamiltonian is
\begin{equation}
H_{\omega 0} = \frac{\epsilon_0}{2}\int d^3\mathbf{r} \left[\mathbf{E}_{\perp}^2(\mathbf{r})+ c^2\mathbf{B}_{\perp}^2(\mathbf{r}) \right]. 
\label{Hradb_0}
\end{equation}
The vector potential is, with $A_0 = \sqrt{\hbar / (2 \epsilon_0 L_c^3)}$,
\begin{equation}
\mathbf{A}(\mathbf{r}) =A_0  \sum_{n,\lambda} \frac{\mathbf{\epsilon}_{n,\lambda}}{\sqrt{\omega_n}} 
\left[ a_{n,\lambda} e^{i\mathbf{k}_n\cdot\mathbf{r}} + a^{\dagger}_{n,\lambda} e^{-i\mathbf{k}_n\cdot\mathbf{r}} \right],
\label{vecpot}
\end{equation}
with $\mathbf{\epsilon}_{n,\lambda}$ the two ($\lambda=1,2$) polarization vectors transversal to the propagation direction indicated by the wavevector $\mathbf{k}_n$, which is related to the allowed frequencies $\omega_n$ in a cavity. The photon annihilation/creation operators  $a_{n,\lambda}$/$a^{\dagger}_{n,\lambda}$  satisfy $[a_{n,\lambda},a^{\dagger}_{n',\lambda'}]=\delta_{n,n'}\delta_{\lambda,\lambda'}$. The Coulomb gauge condition, $\mathbf \nabla \cdot \mathbf A(\mathbf r)=0$, yields the fields
\begin{eqnarray}
\mathbf{E}(\mathbf{r}) &=& i A_0  \sum_{n,\lambda} \sqrt{\omega_n} \mathbf{\epsilon}_{n,\lambda}
\left[ a_{n,\lambda} e^{i\mathbf{k}_n\cdot\mathbf{r}} - a^{\dagger}_{n,\lambda} e^{-i\mathbf{k}_n\cdot\mathbf{r}} \right], \label{EF} \\
\mathbf{B}(\mathbf{r}) &=& i \frac{A_0}{c}  \sum_{n,\lambda} \sqrt{\omega_n} \mathbf{{\kappa}_{n,\lambda}}
\left[ a_{n,\lambda} e^{i\mathbf{k}_n\cdot\mathbf{r}} - a^{\dagger}_{n,\lambda} e^{-i\mathbf{k}_n\cdot\mathbf{r}} \right], \label{BF}
\end{eqnarray}
with $\mathbf{\kappa}_{n,\lambda}=\mathbf k_n \cross \mathbf{\epsilon}_{n,\lambda}/|\mathbf{k}_n|$, so that $H_{\omega 0}$ in \eqref{Hradb_0} becomes
\begin{eqnarray}
H_{\omega 0} &=&  \sum_{n,\lambda} \hbar \omega_n \left[a^{\dagger}_{n,\lambda} a_{n,\lambda} + \frac{1}{2} \right] \label{Hw_op} \\
                     &=& \sum_{n,\lambda} \frac{\hbar \omega_n}{2} \left[q^2_{n,\lambda} - \frac{\partial^2}{\partial q^2_{n,\lambda} } \right]; \label{Hw_dv}
\end{eqnarray}
equation \eqref{Hw_dv} comes after rewriting the field operators in \eqref{Hw_op} as function of the adimensional displacement coordinates $q_{n,\lambda}$ and their conjugate momenta $\partial / \partial q_{n,\lambda}$,
\begin{equation}
a_{n,\lambda} = \frac{1}{\sqrt{2}} \left(  q_{n,\lambda} + \frac{\partial}{\partial q_{n,\lambda}}\right),  a^{\dagger}_{n,\lambda} = \frac{1}{\sqrt{2}} \left(  q_{n,\lambda}  - \frac{\partial}{\partial  q_{n,\lambda}}\right).
\label{aad}
\end{equation}

Assuming the long wavelength approximation, $\mathbf{k}_n \cdot \mathbf{r} \ll 1$, so that $\mathbf{A}=\mathbf{A}(\mathbf{r_0})$ is constant in the device, and by choosing $\mathbf{r_0} = \mathbf{0}$, equation \eqref{vecpot} reads
\begin{eqnarray}
\mathbf{A} &=& A_0  \sum_{n,\lambda} \frac{\mathbf{\epsilon}_{n,\lambda}}{\sqrt{\omega_n}} \left[ a_{n,\lambda} + a^{\dagger}_{n,\lambda}  \right], \label{vecpotdip} \\
                   &=& \sqrt{2} A_0  \sum_{n,\lambda} \frac{\mathbf{\epsilon}_{n,\lambda}}{\sqrt{\omega_n}} q_{n,\lambda}, \label{vecpotzz} 
\end{eqnarray}
in which it is implicit that the sum should include only $M$ modes satisfying $\mathbf{k}_n \cdot \mathbf{r} \ll 1$. For getting a simplified Hamiltonian in the Coulomb gauge we consider here a single mode $\omega$, so that $\mathbf{A}$ reads
\begin{eqnarray}
\mathbf{A} &=& \frac{A_0}{\sqrt{\omega}} \sum_{\lambda} \mathbf{\epsilon}_{\lambda} \left( a_{\lambda} + a^{\dagger}_{\lambda} \right), \label{vecpot_1} \\
                   &=& \sqrt{2} \frac{ A_0}{\sqrt{\omega}}  \sum_{\lambda} \mathbf{\epsilon}_{\lambda} q_{\lambda} . \label{vecpot_2}
\end{eqnarray}
Since in the Coulomb gauge the first term in \eqref{minimalH} becomes $-\hbar^2 \mathbf{\nabla}^2_j + e^2 \mathbf{A}^2 (\mathbf{r}_j) +  i 2 e \hbar \mathbf{A} (\mathbf{r}_j) \cdot \mathbf{\nabla}_j$, the initial Hamiltonian $H = H_0 + H_{\omega 0}$ can be rewritten as $H^{(C)}=H_e + H_{e\omega} + H_\omega$, where the purely matter term is
\begin{equation}
H_e =  \sum_{j=1}^N \left[ -\frac{\hbar^2}{2m_e} \mathbf{\nabla}^2_j + V(\mathbf{r}_j) + eU(\mathbf{r}_j) \right],
\label{H_eN}
\end{equation}
the light-matter interaction term is, by defining the coupling constant $g^{(c)}=e\hbar/m_e\sqrt{\hbar/(2\epsilon_0 L_c^3 \omega)}$,
\begin{eqnarray}
H_{e\omega} &=&  ig^{(c)} \sum_{\lambda} \left( a_{\lambda} + a^{\dagger}_{\lambda} \right) \mathbf{\epsilon}_{\lambda} \cdot \mathbf{\nabla}, \label{H_ew1} \\
                       &=&  i \sqrt{2} g^{(c)} \sum_{\lambda} q_{\lambda} \mathbf{\epsilon}_{\lambda} \cdot \mathbf{\nabla}, \label{H_ew2}
\end{eqnarray}
where $\mathbf{\nabla} = \sum_{j=1}^N \mathbf{\nabla}_j$, and \eqref{vecpot_1} used in \eqref{H_ew1} while \eqref{vecpot_2} used in \eqref{H_ew2}, and the purely field term is, by defining the plasmon frequency $\omega_p=\sqrt{n_e e^2/(\epsilon_0 m_e)}$ with $n_e=N/L_c^3$ the full electron density,
\begin{eqnarray}
H_{\omega} &= &  \hbar \omega \sum_{\lambda} \left[ \frac{\omega_p^2}{4 \omega^2} \left(a_{\lambda} + a^{\dagger}_{\lambda}\right)^2 + \left(a^{\dagger}_{\lambda} a_{\lambda} + \frac{1}{2} \right) \right], \label{H_w1} \\
                    & = &   \frac{\hbar \omega}{2} \sum_{\lambda}  \left[\frac{\omega_p^2}{\omega^2} q^2_{\lambda} +  \left(q^2_{\lambda} - \frac{\partial^2}{\partial q^2_{\lambda} } \right) \right], \label{H_w2}
\end{eqnarray}
with \eqref{Hw_op} and \eqref{vecpot_1} used in \eqref{H_w1}, while \eqref{Hw_dv} and \eqref{vecpot_2} used in \eqref{H_w2}. The orthogonality of $\lambda$ for a single mode renders the simple quadratic term in \eqref{H_w1}, otherwise, crossed terms among different modes would arise; also, the coupling constant $g^{(c)}$ would be different for different modes. Expressions \eqref{H_eN}, \eqref{H_ew1}, \eqref{H_w1} (or \eqref{H_eN}, \eqref{H_ew2}, \eqref{H_w2}) yield the full Hamiltonian $H^{(C)}$ in the Coulomb gauge, which substitute the original expressions in \eqref{minimalH}, \eqref{Hw_op} (or \eqref{minimalH}, \eqref{Hw_dv}). 

The Hamiltonian $H^{(C)}$ could be separated in one purely electronic part, one purely photonic part including the ``$\mathbf{A}^2$'' diamagnetic term, and one ``$\mathbf{A} \cdot \mathbf{P} $'' coupling part. But, in fact, such only is the case when the diamagnetic term is neglected; if included, as \eqref{H_w1} or \eqref{H_w2} shows, the plasmon frequency already indicates the intrinsic polaritonic nature of the problem since the photon Hamiltonian depends on matter parameters. Only when the coupling constant $g^{(c)}$ is weak one may neglect the diamagnetic term but, at stronger couplings, such a term is usually taken care by new `polaritonic' operators \cite{QEDFTno_refereeNOcurrent_2022},  
\begin{eqnarray}
b_\lambda                &=& \frac{1}{2\sqrt{\omega \tilde\omega}} [a_\lambda (\tilde\omega + \omega) + a^\dagger_\lambda (\tilde\omega - \omega) ], \\
b^\dagger_\lambda &=& \frac{1}{2\sqrt{\omega \tilde\omega}} [a_\lambda (\tilde\omega - \omega) + a^\dagger_\lambda (\tilde\omega + \omega) ],
\label{newops}
\end{eqnarray}
where $\tilde\omega = \sqrt{\omega^2 + \omega^2_p}$ is a dressed frequency for the cavity including the diamagnetic shift, and with $[b_\lambda,b^\dagger_{\lambda'}] = \delta_{\lambda,\lambda'}$. Under such new algebra, the vector potential in \eqref{vecpot_1} is
\begin{equation}
\tilde{\mathbf{A}} = \frac{A_0}{\sqrt{\tilde \omega}} \sum_{\lambda} \mathbf{\epsilon}_{\lambda} \left( b_{\lambda} + b^{\dagger}_{\lambda} \right), 
\label{vecpot_new}
\end{equation}
the photonic part in \eqref{H_w1} simply reads
\begin{equation}
\tilde{H}_{\omega} =   \hbar \tilde \omega \sum_{\lambda} \left( b^{\dagger}_{\lambda} b_{\lambda} + \frac{1}{2} \right) , 
\label{H_w1_new}
\end{equation}
and the coupling part in \eqref {H_ew1} becomes, with $\tilde{g}^{(c)}=e\hbar/m_e\sqrt{\hbar/(2\epsilon_0 L_c^3 \tilde{\omega})}$,
\begin{equation}
\tilde{H}_{e\omega} =  i \tilde{g}^{(c)} \sum_{\lambda}\left( b_{\lambda}+ b^{\dagger}_{\lambda} \right) \mathbf{\epsilon}_{\lambda} \cdot \mathbf{\nabla},
\label{H_ew1_new}
\end{equation}
while the electronic part in \eqref{H_eN} remains the same. As such the new full Hamiltonian reads $\tilde H^{(C)} = H_e + \tilde{H}_{e\omega} + \tilde{H}_\omega$, so that the diamagnetic term is embodied in the transformation itself; such a simple form for $\tilde H^{(C)}$ would not be possible if more modes had been included \cite{QEDFTno_refereeNOcurrent_2022}. One could also similarly write $\tilde H^{(C)}$ in terms of new operators for $q_\lambda , \partial /\partial q_\lambda$, but we do not proceed further since, for the evolution equations of our transport model, we opted in working in the Dipole gauge as it will be explained in the end of next subsection.

\subsection{Dipole gauge Hamiltonian}
\label{Hdipole}

The starting point for the gauge transformation are the Hamiltonians $H_0$ in \eqref{minimalH} and $H_{\omega 0}$ in \eqref{Hw_op}, and the vector potential already in dipole approximation in \eqref{vecpotdip}. We will justify the use of the dipolar approximation in our particular case in Sec. \ref{dis}. We remind that, in principle, one should make the gauge transformation \textit{before} the dipole approximation, in the multipole PZW spirit. However, we proceed by applying it \textit{after}, in our goal of an initial qualitative model. Such transformation is given by \cite{PZW_deriv,QEDFT_noGS_2018,cohen1997photon}
\begin{equation}
Z = \text{exp}\left[  - \frac{i e}{\hbar} \sum_{j=1}^N \mathbf{r}_j \cdot \mathbf{A} \right] = \text{exp}\left[  - \frac{i}{\hbar} \mathbf{d} \cdot \mathbf{A} \right],
\label{UT0}
\end{equation}
with $\mathbf{d} = e \mathbf{R}$, $\mathbf{R} = \sum_{j=1}^N \mathbf{r}_j$, the total dipole moment. One needs to calculate the new Hamiltonian $H'$,
\begin{equation}
H' = Z H Z^\dagger + i \hbar \frac{dZ}{dt} Z^\dagger,
\label{Hprime}
\end{equation}
where the second term plays a role only if $\mathbf{A}$ is explicitly time dependent. From \eqref{vecpotdip} and by defining $\chi_{n,\lambda} = i / \sqrt{ 2 \epsilon_0 \hbar \omega_n L_c^3 } \mathbf{\epsilon}_{n,\lambda} \cdot \mathbf{d}$, $Z$ can be recast into
\begin{equation}
Z = \text{exp}\left[ \sum_{n,\lambda} ( \chi^*_{n,\lambda} a_{n,\lambda} - \chi_{n,\lambda} a^\dagger_{n,\lambda} ) \right].
\label{UT}
\end{equation}
The transformation of $H_0$, $H'_0 = Z H_0 Z^\dagger$, yields simply $H_0'=H_e$, the same Hamiltonian shown in \eqref{H_eN}, while the transformation of $H_{\omega 0}$, $H'_{\omega 0} = Z H_{\omega 0} Z^\dagger$, yields in addition to \eqref{Hw_op}, two extra terms,
\begin{equation}
H'_{\omega 0} = H_{\omega 0} + \sum_{n,\lambda} \hbar \omega_n \left[( \chi_{n,\lambda} a^\dagger_{n,\lambda} + \chi^*_{n,\lambda} a_{n,\lambda} )+  \chi^*_{n,\lambda} \chi_{n,\lambda} \right].
\label{extra}
\end{equation}
By noticing from \eqref{EF} that the electric field in dipole approximation is
\begin{eqnarray}
\mathbf{E} &=& i A_0    \sum_{n,\lambda} \sqrt{\omega_n} \mathbf{\epsilon}_{n,\lambda} \left[ a_{n,\lambda} - a^{\dagger}_{n,\lambda} \right], \label{EFdip} \\
                   &=& i \sqrt{2} A_0  \sum_{n,\lambda} \sqrt{\omega_n} \mathbf{\epsilon}_{n,\lambda} \frac{\partial}{\partial q_{n,\lambda}}, \label{EFdip2}
\end{eqnarray}
the extra $\chi$-linear terms in \eqref{extra} yield the so-called ``$\mathbf{E} \cdot \mathbf{R}$'' coupling term, 
\begin{equation}
H_{er} =  -  \mathbf{d}  \cdot \mathbf{E},
\label{H_er}
\end{equation}
while the extra $\chi$-quadratic term in \eqref{extra} yields the dipole self-energy, a parabolic energy as
\begin{equation}
\epsilon_{dip} = \frac{1} {2 \epsilon_0 L_c^3} \sum_{n,\lambda} (\mathbf{\epsilon}_{n,\lambda} \cdot \mathbf{d})^2 = \sum_{n,\lambda} \hbar \omega_n |\chi_{n,\lambda}|^2 . 
\label{Edip}
\end{equation}
Altogether, the full new Hamiltonian in the Dipole gauge can then be cast, with the exchange $q_{n,\lambda} \rightarrow -i \partial / \partial q_{n,\lambda}$ and $i \partial / \partial q_{n,\lambda} \rightarrow q_{n,\lambda}$ that preserves the commutation relations, into $H^{(D)}= H_e + \epsilon_{dip} + H_{\omega 0} + H_{er}$, which in terms of $a_{n,\lambda}$, $a^\dagger_{n,\lambda}$ becomes
\begin{eqnarray}
&&H^{(D)} =  \sum_{j=1}^N \left[ -\frac{\hbar^2  \mathbf{\nabla}^2_j}{2m_e} + V(\mathbf{r}_j) + eU(\mathbf{r}_j)\right] + \sum_{n,\lambda} \hbar \omega_n |\chi_{n,\lambda}|^2  \nonumber \\
   &+&   \sum_{n,\lambda} \left(\hbar \omega_n \left[a^{\dagger}_{n,\lambda} a_{n,\lambda} + \frac{1}{2} \right] +  \hbar \omega_n |\chi_{n,\lambda}| \left[ a_{n,\lambda} + a^\dagger_{n,\lambda} \right] \right),
\label{HdipOP}
\end{eqnarray}
where \eqref{Hw_op}, \eqref{EFdip}, \eqref{H_er} were used, while in terms of $q_{n,\lambda}$, $\partial / \partial q_{n,\lambda}$ it becomes
\begin{eqnarray}
&&H^{(D)} =  \sum_{j=1}^N \left[ -\frac{\hbar^2  \mathbf{\nabla}^2_j}{2m_e} + V(\mathbf{r}_j) + eU(\mathbf{r}_j)\right] + \sum_{n,\lambda} \hbar \omega_n |\chi_{n,\lambda}|^2 \nonumber \\
   &+& \sum_{n,\lambda} \left(\frac{\hbar \omega_n}{2} \left[q^2_{n,\lambda} - \frac{\partial^2}{\partial q^2_{n,\lambda} } \right] + \sqrt{2} \hbar \omega_n |\chi_{n,\lambda}| q_{n,\lambda} \right),
 \label{Hdip}
\end{eqnarray}
where \eqref{Hw_dv}, \eqref{EFdip2}, \eqref{H_er} were used. It can also be cast into
\begin{eqnarray}
H^{(D)} &=&  \sum_{j=1}^N \left[ -\frac{\hbar^2}{2m_e} \mathbf{\nabla}^2_j + V(\mathbf{r}_j) + eU(\mathbf{r}_j)\right] \label{Hdip2} \\
   &+ &  \sum_{n,\lambda} \frac{\hbar \omega_n}{2} \left[ - \frac{\partial^2}{\partial q^2_{n,\lambda} } + \left( q_{n,\lambda} - \sqrt{2} |\chi_{n,\lambda}| \right)^2 \right], \nonumber
\end{eqnarray}
with the second line a shifted harmonic oscillator \cite{QEDFT_noGS_2018}. Equation \eqref{Hdip} is the Bohmian starting point in Sec. \ref{Hbohm}.

Three important features arise when comparing Coulomb and Dipole gauges. First, the diamagnetic term is either neglected or included via a new algebra in the Coulomb gauge, while in the Dipole gauge it is absorbed by the gauge transformation itself; the squared $R^2$ term in $\epsilon_{dip}$ has no resemblance to the $A^2$ term. Second, the intrinsic polaritonic nature of the problem, embodied in $\omega_p$ in the Coulomb gauge, in the Dipole gauge is embodied by $\epsilon_{dip}$, which makes possible to go from \eqref{Hdip} to \eqref{Hdip2} showing how the photonic part is written as a shifted harmonic oscillator dictated by the electron dipole in $\chi_{n,\lambda}$; so one cannot think of purely electronic and photonic terms in $H^{(D)}$, for example, by identifying the purely electronic part as $H_e + \epsilon_{dip}$. The polaritonic nature of the coupled light-matter system is evident, since the photon coordinate is related to a displacement field containing both electric field and matter polarization. Third, different observables have different meanings in different gauges \cite{QEDFT_refereePZW_2020,cohen1997photon}. For example, the velocity is $\mathbf{v} = (\mathbf{p} + e \mathbf{A})/m_e$, while $\mathbf{v'} = Z \mathbf{v} Z^{\dagger}= \mathbf{p}/m_e$, so that the current is also different among the gauges. This will be important when discussing Bohmian velocities in next section. The transversal electric field also differs since $\mathbf{E'}_{\perp} = Z \mathbf{E}_{\perp} Z^{\dagger}= \mathbf{E}_\perp - \mathbf{P}_\perp/\epsilon_0$, with the transversal polarization $\mathbf{P}_\perp = 1/(2 \pi \L_c^3) \sum_{n,\lambda} [\mathbf{\epsilon}_{n,\lambda} (\mathbf{\epsilon}_{n,\lambda} \cdot \mathbf{d}) e^{i\mathbf{k}_n\cdot\mathbf{r}} + \textit{c.c.}]$; this is related to the fact that, in the Dipole gauge, it is trully the transversal displacement field $\mathbf{D}_\perp = \epsilon_0  \mathbf{E}_\perp +  \mathbf{P}_\perp$ that it is quantized, although in our transport model, with no bound charges but only tunneling electrons, such a formal difference among $\mathbf{D}$ and $\mathbf{E}$ should play no role; also, in the full PZW multipole spirit, it is $\mathbf{P}$ instead of $\mathbf{d}$ that would define the unitary transformation in \eqref{UT0}.

\subsection{Semiclassical Hamiltonian}
\label{Hsemi}

The `standard' approach for a semiclassical treatment of the problem, where the electromagnetic field is no longer quantized so that no quantum entanglement between matter and light is considered, is to take $ q_{n,\lambda} \rightarrow  q_{n,\lambda}(t)$ so that potentials and fields are treated as time-dependent. As such, when we apply into \eqref{minimalH} and \eqref{Hw_dv} the transformation \eqref{Hprime}, but this time with a time-dependent $\mathbf{A}(t)$ in \eqref{vecpotzz} we get, in addition to \eqref{Hdip}, the extra term $ i \hbar (dZ / dt) Z^\dagger = -\mathbf{d} \cdot \mathbf{E}(t)$, since $ \mathbf{E}(t) = - \partial \mathbf{A}(t) / \partial t$. So, when the semiclassical limit is taken \textit{after} the gauge transformation, its Hamiltonian $H^{(SC)}(t)$ reads \cite{cohen1997photon}
\begin{equation}
H^{(SC)}(t) =  H_e  - \mathbf{d} \cdot \mathbf{E}(t) + \epsilon_{dip},
\label{HSC2} 
\end{equation}
with $H_e$ from \eqref{H_eN}, and where the purely photonic part in the second line of \eqref{Hdip} is not included since it becomes only a time-dependent potential, with no spatial dependence and no effect on the dynamics.

Notice that the semi-classical approach, under a self-consistent treatment of electron transport and classical electromagnetic fields, still includes a classical back-action between electron dynamics and fields. A more fundamental way of mixing the degrees of freedom $\vec r_j$ and  $q_{n,\lambda}$ with the time-dependent functions $\vec r_j(t)$ and $q_{n,\lambda}(t)$, without losing any fundamental rigour in the development of light-matter coupling models, with quantum and classical back-actions, is approached via Bohmian formalism in the next section.  

\section{Displacement current in nanodevices}
\label{dis}

We model in this section the total (particle plus displacement) current for nanodevices at THz frequencies. We use Bohmian theory to develop an expression for the total current in terms of electronic velocities, which we will use later in Sec. \ref{res} to access information on the polaritonic nature of the system. A displacement current coefficient - instead of a static transmission coefficient - is required in AC scenarios with operation frequencies comparable to the inverse of the electron dwell time. 

\subsection{Maxwell equations}
\label{ME}

In terms of $3$D electric $\mathbf E(\mathbf r,t)$ and magnetic $\mathbf B(\mathbf r,t)$ fields in vacuum, from Helmholtz decomposition, the Maxwell equations can be written \cite{cohen1997photon, grynberg, scully} for the transverse electric field as
\begin{equation}
\mathbf \nabla \cdot \mathbf E_{\perp}(\mathbf r,t)=0, \;\;\; \mathbf \nabla \times \mathbf E_{\perp}(\mathbf r,t)=-\frac{\partial \mathbf B(\mathbf r,t)}{\partial t}, 
\label{faraday}
\end{equation}
for the longitudinal electrical field as 
\begin{equation}
\mathbf \nabla \times \mathbf E_{\parallel}(\mathbf r,t)=0, \;\;\; \mathbf \nabla \cdot \mathbf E_{\parallel}(\mathbf r,t)=\frac{\rho(\mathbf r,t)}{\epsilon_0}, 
\label{gauss}
\end{equation}
and for the transverse ($\mathbf{B}_{\parallel}=0$) magnetic field as 
\begin{equation}
\mathbf \nabla \cdot\mathbf B_{\perp}(\mathbf r,t)=0, \;\;\; \mathbf \nabla \times \mathbf B_{\perp}(\mathbf r,t)=\mu_0 \epsilon_0\frac{\partial \mathbf E(\mathbf r,t)}{\partial t}+\mu_0 \mathbf J_c(\mathbf r,t).
\label{amper}
\end{equation}
In such expressions, longitudinal and transverse components are coupled through the charge density
\begin{equation}
\rho(\mathbf r,t)=\sum_{j=1}^N e \delta(\mathbf r-\mathbf r_j[t]),
\label{rho}
\end{equation}
and the particle current density
\begin{equation}
\mathbf J_c(\mathbf r,t)=\sum_{j=1}^N e \mathbf v_j(t) \delta(\mathbf r-\mathbf r_j[t]).
\label{J}
\end{equation}
We mention that \eqref{rho} and \eqref{J} remain valid in Bohmian quantum mechanics \cite{Xavier_book,Bohm1,Bohm2,Xavier_Ferry}, where each particle has well-defined position $\mathbf r_j(\mathbf r,t) $ and velocity $\mathbf v_j(\mathbf r,t)$. From the divergence of the Ampère-Maxwell equation, we reach the definiton of the total current $\mathbf J(\mathbf r,t)$ as the sum of three contributions,
\begin{equation}
\mathbf J(\mathbf r,t)=\mathbf J_c(\mathbf r,t)+\epsilon_0\frac{\partial \mathbf E_{\parallel}(\mathbf r,t)}{\partial t}+\epsilon_0\frac{\partial \mathbf E_{\perp}(\mathbf r,t)}{\partial t},
\label{total}
\end{equation}
in which the last two terms yield the displacement current density, so that $\nabla \cdot \mathbf J(\mathbf r,t)=0$ is satisfied to ensure that the total current $I(t)$, predicted on a surface of the active region of an electron device, is equal to the total current $I'(t)$ measured on a surface of an ammeter as in Fig. \ref{volume}(c), at least when radiation outside the connecting cables is assumed negligible. In other words, at high frequency, the particle current $\mathbf J_c(\mathbf r,t)$ on the surface of the active region is not equal to the particle current $\mathbf J_c(\mathbf{r} _M,t)$ on a surface of the ammeter at the position $\mathbf{r} _M$. Thus, in high-frequency electronics, what has to be predicted is the total current in the active region, $\mathbf J(\mathbf r,t)$, and not only the particle current $\mathbf J_c(\mathbf r,t)$. 

\subsection{Bohmian formalism}
\label{Hbohm}

The direct modelling of \eqref{total} can be done, in principle, from both longitudinal and transversal fields in \eqref{faraday}, \eqref{gauss}, \eqref{amper} self-consistently solved with \eqref{rho}, \eqref{J}. However, in our work we consider a different approach (see next subsection), where \eqref{total} is instead computed from the velocities of the electrons; it has the advantage of evidencing a direct link between electron dynamics and total current, that will prove useful when interpreting results in Section \ref{results_D_AC}. The well-known Ramo-Shockley-Pellegrini theorem \cite{Ramo,Shockley,Pellegrini_1986} shows the equivalence between the two methods. For quantum systems, the velocity of the electrons can be obtained from the Bohmian theory \cite{Damiano}, as described in this subsection.

Our starting equation is the Dipole Hamiltonian $H^{(D)}$ in \eqref{Hdip}, in position representation, where $\mathbf{r}_j=\{x_j,y_j,z_j\}$ is interpreted as the $j$-electron position and $q_{n,\lambda}$ as the amplitude of the $n,\lambda$-mode of the field. The many-electron many-mode wave function is then
\begin{equation}
\Psi(\mathbf{W},t) \equiv \Psi(x_1,y_1,...,z_N,q_{1,1},q_{1,2},...,q_{M,2},t),
\end{equation}
where $\mathbf{W}=\{x_1,y_1,..,z_N,q_{1,1},q_{1,2},..,q_{M,2}\}$ is a point in the configuration space of $N$ electrons and $2M$ modes. Such wave function is solution of the time-dependent many-body Schroedinger equation
\begin{equation}
i\hbar \frac{\partial}{\partial t} \Psi(\mathbf{W},t)=H^{(D)} \Psi(\mathbf{W},t),
\end{equation}  
which allows us to develop a \textit{local-in-position} continuity equation for the presence probability,
\begin{equation}
i\hbar \frac{\partial |\Psi(\mathbf{W},t)|^2}{\partial t} + \sum_{j} \nabla_j \mathbf J_j(\mathbf{W},t)+\sum_{n,\lambda} \frac{\partial J_{n,\lambda}(\mathbf{W},t)}{\partial q_{n,\lambda}} =0,
\label{conti}
\end{equation}  
where
\begin{eqnarray}
\mathbf J_j(\mathbf{W},t)&=&i\frac{\hbar}{2m} \left(\Psi \nabla_j \Psi^*-\Psi^*\nabla_j \Psi\right), \label{jr} \\
J_{n,\lambda}(\mathbf{W},t)&=&i\frac{\omega_n}{2} \left(\Psi \frac{\partial \Psi^*}{\partial q_{n,\lambda}}-\Psi^*\frac{\partial \Psi}{\partial q_{n,\lambda}}\right). \label{jq}
\end{eqnarray}
According to Bohmian theory, the continuity equation \eqref{conti} allows us to define a velocity field for each electron,
\begin{equation}
v_{x_j}(\mathbf{W},t)=\frac{J_{j,x}(\mathbf{W},t)}{|\Psi(\mathbf{W},t)|^2}=\frac{1}{m_e}\frac{\partial S(\mathbf{W},t)}{\partial x_j},
\label{vx}
\end{equation}   
where $J_{j,x}(\mathbf{W},t)$ is the $x_j$ component of $\mathbf J_j(\mathbf{W},t)$; the velocity fields $v_{y_j}(\mathbf{W},t)$ and $v_{z_j}(\mathbf{W},t)$ are computed identically. A ``velocity" field for each amplitude of the mode $q_{n,\lambda}$ can be computed as 
\begin{equation}
v_{n,\lambda}(\mathbf{W},t)=\frac{J_{n,\lambda}(\mathbf{W},t)}{|\Psi(\mathbf{W},t)|^2}=\frac{\omega_n}{\hbar}\frac{\partial S(\mathbf{W},t)}{\partial q_{n,\lambda}}.
\label{vq}
\end{equation}   
In these expressions we have rewritten the complex wave function in polar form, $\Psi(\mathbf{W},t)=R(\mathbf{W},t)e^{iS(\mathbf{W},t)/\hbar}$, with $R(\mathbf{W},t)$ and $S(\mathbf{W},t)$ real functions. From such velocities one obtains the set of coupled trajectories
\begin{eqnarray}
x^l_{j}(t)&=&x^l_{j}(0)+\int_0^t  v_{x_j}(\mathbf{W}(t'),t') dt', \nonumber \\
q^l_{n,\lambda}(t)&=&q^l_{n,\lambda}(0)+\int_0^t v_{n,\lambda}(\mathbf{W}(t'),t') dt',
\label{ap_traj}
\end{eqnarray}
where $\mathbf{W}(t')$ are the Bohmian trajectories of all degrees of freedom. The superindex $l$ specifies the initial variables $x^l_{j}(0)$ and $q^l_{n,\lambda}(0)$, different in each experiment $l$; the ``trajectory" for $q^l_{n,\lambda}(t)$ is to be taken as a purely mathematical tool. Since all experiments are considered to be described by the same identically prepared wave function $\Psi(\mathbf{W},t)$, the distribution of the different $x^l_{j}(0)$ and $q^l_{n,\lambda}(0)$ and of all others degrees of freedom has to be consistent with $|\Psi(\mathbf{W},t)|^2$. It can be demonstrated that for a large number of experiments $N_{\text{exp}}$, at any time $t$ the identity \cite{Bohm1}
\begin{eqnarray}
|\Psi(\mathbf{W},t)|^2=\lim_{N_{\text{exp}}\to\infty} \sum_{l=1}^{N_{\text{exp}}} \displaystyle \prod_{s} \delta (w-w^l_s(t))
\label{sumc}
\end{eqnarray}
is satisfied, where $s$ runs for all degrees of freedom of $\mathbf W$. These Bohmian velocities will later be used to provide an intrinsic value of the total current \cite{Xavier_PRL,Norsen,Devashish,Damiano,Eisenberg,Eisenberg2}.  

Since the Bohmian formalism handles both the degrees of freedom in $\mathbf W$ and their related trajectories $\mathbf W(t)$, we can construct the so-called Bohmian conditional wave function by mixing such two features. By defining $\mathbf{W}_j=\{x_1,y_1,z_1,...,y_j,z_j,...,z_N\}$ as all electron degrees of freedom except $x_j$, and $\mathbf{W}_q=\{q_{1,1},q_{1,2},...,q_{M,1},q_{M,2}\}$ as all degrees of freedom of the field, we can obtain a set of $N$ conditional Bohmian wave packets
\begin{eqnarray}
\psi^{1}(x_1,\mathbf{W}_q,t) &\equiv& \Psi(x_1,\mathbf W_q,\mathbf{W}_1(t),t), \nonumber\\
&....&\nonumber\\
\psi^{j}(x_j,\mathbf{W}_q,t) &\equiv& \Psi(x_j,\mathbf W_q,\mathbf{W}_j(t),t), \nonumber\\
&....&
\label{flux}
\end{eqnarray}
such that $\mathbf{W}=\{x_j,\mathbf{W}_j,\mathbf{W}_q\}$. It is straightforward to see that the velocities in \eqref{vx} can be computed equivalently from either $\psi^{j}(x_j,\mathbf{W}_q,t)$ or $\Psi(\mathbf{W},t)$ \cite{Xavier_book,Xavier_PRL}. From the computational side, the advantage in dealing with a set of wave packets $\psi^{j}(x_j,\mathbf{W}_q,t)$  is that they dwell in a much smaller configuration space. In fact, $\psi^{j}(x_j,\mathbf{W}_q,t)$ can be understood as a pure state description of an open system (instead of a reduced density matrix) for either Markovian or non-Markovian scenarios \cite{Wiseman,Wiseman2}. Therefore, we can recover the picture of transport as a flux of $j=1,..,N$ wave-packets $\psi^{j}(x_j,\mathbf{W}_q,t)$ \cite{Xavier_book,Xavier_PRL}. We will later use such conditional states. 

\subsection{The Ramo-Shockley-Pellegrini theorem}
\label{ram}

The computation of the total current follows from the Ramo-Shockley-Pellegrini theorem \cite{Ramo,Shockley,Pellegrini_1986}. We consider an arbitrary volume $\Omega$ with length $L_{\Omega}$, height $H_{\Omega}$, and width $W_{\Omega}$, limited by a closed surface $\mathcal{S}$ which, in turn, is composed of six rectangular surfaces $\mathcal{S}=\{\mathcal{S}_1,...,\mathcal{S}_6\}$. Notice that $\Omega$ is just an arbitrary mathematical volume used in the computation of the displacement current (much larger than the volumes in Fig \ref{volume}(a)). The total current $I_{\eta}(t)=\int_{\mathcal{S}_{\eta}} \mathbf J(\mathbf x,t)\cdot d\mathbf s$ is computed on one of the surfaces of $\mathcal{S}$ named $\mathcal{S}_{\eta}$. We define a set of functions (one for each particular surface $\mathcal{S}_{\eta}$) as scalar functions $G_{\eta}(\mathbf r)$ and vector functions $\mathbf F_{\eta}(\mathbf r)$ through
\begin {equation}
\mathbf F_{\eta}(\mathbf r)=-\mathbf \nabla G_{\eta}(\mathbf r),
\label{equation_3}
\end {equation}
and fulfilling
\begin {equation}
\mathbf \nabla \cdot \left[ \epsilon_0 \mathbf F_{\eta}(\mathbf r)\right]=-\mathbf \nabla \cdot \left[ \epsilon_0 \mathbf \nabla G_{\eta}(\mathbf r)\right]=0.
\label{equation_4}
\end {equation}
For reasons that will be later clear, the following particular Dirichlet boundary conditions on the shape of $G_{\eta}(\mathbf r)$,
\begin {equation}
 G_{\eta}(\mathbf r)=1\;\;\;\forall \mathbf r \in \mathcal{S}_{\eta}  \textrm{   and   }  G_{\eta}(\mathbf r)=0\;\;\;\forall \mathbf r \in \mathcal{S}_{\eta' \neq \eta},
\label{equation_5}
\end {equation}
are considered, meaning that $G_{\eta}(\mathbf r)=1$ on the surface $\mathcal{S}_{\eta}$  and zero on the other surfaces. With \eqref{equation_4} and \eqref{equation_5}, the total current $I_{\eta}(t)$ on $\mathcal{S}_{\eta}$ can be written as
\begin {eqnarray}
 I_{\eta}(t)=\int_\mathcal{S} G_{\eta}(\mathbf r) \mathbf J(\mathbf r,t)\cdot d\mathbf s=-\int_{\Omega}\mathbf F_{\eta}(\mathbf r)\cdot \mathbf J(\mathbf r,t) dv.
\label{equation_6}
\end {eqnarray}
From the total current density in \eqref{total}, and since $\mathbf E_{\parallel}(\mathbf r,t)=- \nabla U(\mathbf r ,t)$, $\mathbf E_{\perp}(\mathbf r,t)=-\partial \mathbf A(\mathbf r ,t)/\partial t$, we get
\begin {eqnarray}
 I_{\eta}(t)&=& -\int_{\Omega}\mathbf F_{\eta}(\mathbf r)\cdot \mathbf J_c(\mathbf r,t) dv\nonumber\\
 &+&\epsilon_0 \int_{\Omega}\mathbf F_{\eta}(\mathbf r) \cdot  \frac{\partial \mathbf \nabla U(\mathbf r ,t)}{\partial t} dv \nonumber\\
 &+&\epsilon_0 \int_{\Omega}\mathbf F_{\eta}(\mathbf r) \cdot  \frac{\partial^2 \mathbf A(\mathbf r ,t)}{\partial t^2} dv.
\label{equation_7}
\end {eqnarray}
This result is a well-known expression found by Ramo \cite{Ramo}, Shockley \cite{Shockley}, and later by Pellegrini \cite{Pellegrini_1986}. 

The \textit{first} term in \eqref{equation_7} cannot be only identified with the particle current since it also includes the displacement current. It contains a volume rather than a surface integral. By using \eqref{J} we see that an electron in the middle of the active region (far from the surface $\mathcal{S}_\eta$) has still a contribution to the total current given by $e \; \mathbf v(\mathbf{r}_j(t),t)\cdot \mathbf F_{\eta}(\mathbf r_j(t))$, which cannot be the particle current because the electron, in the middle of the active region, has not crossed the surface $\mathcal{S}_\eta$ yet, so that the contribution to the total current in this case comes fully from the displacement current. From \eqref{J} we obtain 
\begin {eqnarray}
 -\int_{\Omega}\mathbf F_{\eta}(\mathbf r)\cdot \mathbf J_c(\mathbf r,t) dv = \frac{e}{L_x} \sum_{j=1}^N  \mathbf v_j(t)\cdot \mathbf{{x}},
\label{curfinal1}
\end{eqnarray}
where $\mathbf{v}_j(t)$ is the Bohmian velocity of the $j$ particle in our $N$ particle system. Notice that, because of the spatial integral in \eqref{curfinal1} over the volume $\Omega$, the symbol $N$ does only take into account the electrons inside the volume $\Omega$ at time $t$. Now, as before,   we have considered that the lateral dimensions of $\Omega$ are much larger than the transport direction, $L_{\Omega}\ll H_{\Omega},W_{\Omega}$ so that $\mathbf F_{\eta}(\mathbf r)=-\frac{1}{L_x} \mathbf{{x}}$. 

The \textit{second} term in \eqref{equation_7} can be rewritten as
\begin {eqnarray}
\epsilon_0 \int_{\Omega}\mathbf F_{\eta}(\mathbf r) \cdot  \frac{\partial \mathbf \nabla U(\mathbf r ,t)}{\partial t} dv=\epsilon_0 \int_\mathcal{S}  \frac{d U(\mathbf r ,t)}{dt} \mathbf F_{\eta}(\mathbf r)\cdot d\mathbf s.
\label{equation_8}
\end {eqnarray}
Because of $\nabla U(\mathbf r ,t)$ we have been able to change the evaluation of the scalar potential in the volume $\Omega$, into the evaluation of the outer surface $\mathcal S$, which can easily be done by taking such a volume as large enough so that the surfaces  $\mathcal{S}=\{\mathcal{S}_1,... \, ,\mathcal{S}_6\}$ are either touching the metallic reservoirs that connect the active region to the external battery or are located very far from the active region, so that the scalar potential is zero there. When dealing with a DC battery this term vanishes, while for AC battery it is trivially computed as
\begin {eqnarray}
\epsilon_0 \int_\mathcal{S}  \frac{d U(\mathbf r ,t)}{dt} \mathbf F_{\eta}(\mathbf r)\cdot d\mathbf s=C \frac{d U(\mathbf r ,t)}{dt},
\label{equation_8_bis}
\end {eqnarray}
with $C=\int_\mathcal{S}  \mathbf F_{\eta}(\mathbf r)\cdot d\mathbf s=\epsilon_0 A_c/L_x$, and $A_c$ the lateral area of RTD where $U(\mathbf r ,t)$ is non-zero. Once more $L_{\Omega}\ll H_{\Omega},W_{\Omega}$ is assumed. Such a term mimics a trivial capacitance, with no relevance on (polaritonic features of) the displacement current; such irrelevant geometric factors can easily be eliminated from typical de-embedding techniques in high-frequency electronics \cite{dem}.

The \textit{third} term in \eqref{equation_7} is related to the spatial dependence of the vector potential, which takes into account the (quantized) radiation. It is a complicated term either from theoretical or experimental points of view. Its predictions require going beyond the dipole approximation, and the frequency range where it is relevant is typically higher than what is measured in electron devices. We will typically be interested in components of the displacement current below (about $10 \%$ of) the resonant frequency of the optical cavity, such that this term could also be neglected. Typical measuring electronic setups have to be considered as low-pass filters.

\subsection{Displacement current coefficient}
\label{disD}

The transmission coefficient $T$ is the typical link between the microscopic dynamics of the electrons and the macroscopic current in the device. However, it assumes that the potential profile is constant during all the time $\tau$ that the electron takes to traverse the active region (typically a tunneling region). Such approximation is valid as far as the external frequencies $f$ involved in the performance of the device are much smaller than $1/\tau$. At THz frequencies we can no longer assume that $f\ll 1/\tau$ and the link of the current with the transmission coefficient can be misleading. Another way of understanding this relevant point is that, by construction, the transmission coefficient quantifies the number of particles traversing a particular surface, but we have just seen that even when electrons are not traversing one surface a displacement current is present there. 

From the previous discussions about the three terms in \eqref{equation_7}, we can reasonably assume that the measured total current will then be
\begin {eqnarray}
 I^{(f)}(t)= \frac{e}{L_x} \sum_{j=1}^N  \mathbf v_j(t)\cdot \mathbf{{x}}.
\label{curfinal2}
\end{eqnarray}
We have eliminated the dependence on the surface $\eta$ because in a two terminal device the total current in the emitter is equal to the one in the collector at any time. We have added a dependence on the external frequency $f$ of the emitter-collector bias that provides a time-dependent external scalar potential $V(x,t)$. The velocity of each of the electrons in \eqref{curfinal2} is given by \eqref{vx}. 

We consider now the picture of transport as a flux of different wave packets as mentioned in \eqref{flux}. For large enough $N$ the $j$-sum in \eqref{curfinal2} can be translated into an integral over the different parameters identifying each wave packet $\psi^{j}(x_j,\mathbf{W}_q,t)$, like energy $E$ or injection time $t_i$. Each wave packet is injected at time $t=t_i$ around a central position $x_{left}$ located deep inside the left reservoir with a well-defined central energy $E$ (Fig. \ref{volume}(c)); identical procedure is found for injection from the right reservoir. Thus,
\begin{eqnarray}
\sum_{j=1}^{N(t)} \to  \Gamma \int_{-\infty}^{t} dt_i \int_{-L_x/2}^{L_x/2} dx \int d\mathbf{W}_q |\psi(x,\mathbf{W}_q,t;t_i,E)|^2.
\label{int}
\end{eqnarray}
The value $|\psi(x,\mathbf{W}_q,t;t_i,E)|^2$ quantifies the number of electrons with given values $x$ and $\mathbf{W}_q$ in the interval $dx\; d\mathbf{W}_q$  belonging to a unique wave packet (same $t_i$ and $E$). Notice that $x$ is integrated from $-L_x/2$ to $L_x/2$ to account only for those electrons inside the active region. The integral on $t_i$ does not consider wave packets injected at times $t_i>t$ because such electrons will not be present at time $t$ in the active region. For the moment we have disregarded the integral on the central energy $E$ in \eqref{int}, which will be addressed later. Notice that $\Gamma\equiv\Gamma(E)$ is a parameter indicating the number of electrons being injected from the contact into the barrier region per second (related to the lateral area of the device) per energy unit, that is needed to translate the original discrete sum to our continuous integral.  From \eqref{int} we can rewrite \eqref{curfinal2} for each energy as
\begin{equation}
I^{(f)}(E,t)= \frac{e\Gamma}{L} \int_{-\infty}^{t} dt_i \int_{-L_x/2}^{L_x/2}dx \int d\mathbf{W}_q J_x(x,\mathbf{W}_q,t;t_i,E),
\label{curfinal3}
\end{equation} 
where we have used $J_x=|\psi|^2 v_x$ as indicated in \eqref{vx}. It is very important to realize that a wave packet injected at $t=t_i$ is not equal to a wave packet injected at $t=0$ with a time offset, $\psi(x,\mathbf{W}_q ,t;t_i,E)\neq \psi(x,\mathbf{W}_q ,t-t_i;0,E)$. The reason is that we are interested in temporal variations of the potential $V(x,t)$, so that a wave packet injected at $t_i$ or at $0$ will see different time evolutions of $V(x,t)$.  We define the displacement current coefficient $D$ as the quotient between the current $I^{(f)}(t)$ in \eqref{curfinal3} and the current injected from the contact $e\Gamma(E)$, that is
\begin{eqnarray}
&&D^{(f)}(E, t)=\frac{I^{(f)}(E,t)}{e \Gamma} \label{D_definition_60} \\
&=& \frac{1}{L} \int_{-L_x/2}^{L_x/2} dx \int d\mathbf{W}_q  \int_{-\infty}^{t} dt_i J_x(x,\mathbf{W}_q ,t;t_i,E),\nonumber
\end{eqnarray}
which takes the role of the transmission coefficient $T$ for scenarios where $V(x,t)$ varies in time intervals comparable to the electron transit time, that is, when $\psi(x,\mathbf{W}_q ,t;t_i,E)\neq \psi(x,\mathbf{W}_q ,t-t_i;0,E)$. 

When $V(x,t) \approx V(x)$ is time-independent one can assume $\psi(x,\mathbf{W}_q ,t;t_i,E)=\psi(x,\mathbf{W}_q ,t-t_i;0,E)$, and then $J_x(x,\mathbf{W}_q ,t;t_i,E)=J_x(x,\mathbf{W}_q ,t-t_i;0,E)$, so that \eqref{D_definition_60} can be rewritten with a change of variables $t'=t_i-t$ as
\begin{eqnarray}
D^{(f)}(E,t) &=& \frac{1}{L} \int_{-L_x/2}^{L_x/2} dx \int d\mathbf{W}_q  \int_{0}^{\infty} dt' J_x(x,\mathbf{W}_q ,t';0,E)\nonumber\\
      &=& \frac{1}{L} \int_{-L_x/2}^{L_x/2} dx \; T(E)=T(E),
\label{T_definition_2}
\end{eqnarray}
which provides a time-independent displacement current coefficient. The transmission coefficient $T(E)$ is defined, independently of $D^{(f)}(E,t)$, as
\begin{eqnarray}
T(E) &=& \int d\mathbf{W}_q  \int_{0}^{\infty} dt J_x(x,\mathbf{W}_q ,t;0,E)\nonumber\\
&=&\int d\mathbf{W}_q  \int_{x}^{\infty} dx |\psi(x,\mathbf{W}_q ,\infty;0,E)|^2,
\label{T}
\end{eqnarray}
which is  independent on any $x$ in the active region because, at $t=0$, the wave packet was located far from such region and, at $t\to\infty$, it has been either fully transmitted or reflected, with zero probability in the active region. Notice that $T(E)$ does not depend on time, while in general $D^{(f)}(E,t)$ does. In summary, $D^{(f)}(E,t)\equiv T(E)$ in DC; in AC, $D^{(f)}(E,t)$ in \eqref{D_definition_60} is a novelty of our work.

To reach the final result one has to consider all types of wave packets involved in the transport, e.g., one has to include all relevant central energies. From \eqref{D_definition_60} the total current in \eqref{curfinal3}, when taking into account all relevant energies, can be written in a Landauer-like shape as
\begin{eqnarray}
I^{(f)}(t)=e\int_{-\infty}^{\infty}dE \cdot \Gamma(E) \cdot f_{EF}(E) \cdot D^{(f)}(E,t),
\label{I_Fermi}
\end{eqnarray}
where $f_{EF}(E)$ is the Fermi distribution function and $\Gamma(E)$ is the density of states. We consider positive energy for electrons injected from the left, and negative from the right. The superindex $f$ is the external frequency of the potential, which underlines that $D^{(f)}(E,t)$ can be computed both in DC ($f \ll 1/\tau$) or AC ($f \ge 1/\tau$) conditions.

\section{Simplified coupled electron-photon equations}
\label{WB}

In the search for our qualitative model, we develop in this section the simple (ballistic) implementation of electron transport in a RTD inside an optical cavity. For that we consider different electrons (different injecting times and energies) but only one mode $q$ for the field, corresponding to the cavity resonant mode in Fig. \ref{volume}, with polarization in $x$-direction. Thus, the conditional wave functions in \eqref{flux} can be rewritten as $\psi(x,q,t)$. Under the simplifying assumption that the different electrons do not interact among them but only with the optical mode, the Hamiltonian describing the evolution of each electron coupled to the electromagnetic mode is
\begin{equation}
H^{(D)}_{xq} = -\frac{\hbar^2}{2m_e}\frac{\partial^2}{\partial x^2} + V(x) - \frac{\hbar \omega}{2} \frac{\partial^2}{\partial q^2 } + \frac{\hbar \omega}{2} q^2  + \sqrt{2} \alpha q x,
\label{Hfull}
\end{equation}
where the coupling constant $\alpha = \sqrt{e^2 \hbar \omega N /(2 \epsilon_0 L_c^3)}$ has dimension of energy/length, since $q$ is adimensional. Appendix \ref{bohmderiv} details the steps in going from the original many-electron many-mode Hamiltonian in \eqref{Hdip} to the effective Hamiltonian in \eqref{Hfull}, which resembles the exact Schroedinger equation for a single electron $x$ interacting with a single mode $q$. In Appendix \ref{bohmderiv} we also discuss the path to introduce some of the correlation effects from \eqref{Hdip} neglected in \eqref{Hfull}; in particular, we address how the coupling constant $\alpha$ can account for the fact that $N$ electrons simultaneously interact with the single mode. On that topic, we notice that $\alpha L_x q (x/L_x) \equiv \hbar \omega_r q (x/L_x)$, yielding the Rabi frequency $\omega_r=\alpha L_x /\hbar$. If one then defines the adimensional `oscillator strength' $\beta=2 m_e \omega L_x^2/\hbar$ ($L_x$ takes the role of the dipole matrix element $\langle 0 | x | 1 \rangle$ in these estimations), one can also rewrite  $\alpha L_x q (x/L_x) = \hbar \omega_p \sqrt{\beta}/2 q (x/L_x)$, resembling what found elsewhere \cite{ISB_ultrastrong_THz,RTD_cavityexp}; that is, $\omega_r \equiv \omega_p \sqrt{\beta}/2$, so that the Rabi frequency is dictated by the plasmon frequency of the $N$-electron system. Since $\alpha L_x$ depends on both material and cavity parameters $N$, $L_x$, $L_c$, the coupling constant can be experimentaly engineered as to reach from weak to strong coupling regimes as typically done in experiments \cite{ISB_THz,ISB_ultrastrong_THz,RTD_cavityexp,RTD_polaritonexp} over different platforms. 

The state of the system $\psi(x,q,t)$ is evolved as a $2$D Schroedinger equation from \eqref{Hfull}, whose numerical solution is our goal. For such a goal we will convert it into two coupled $1$D Schroedinger equations. One can define the single-electron eigenstates $\phi_n^{(e)}(x)$, $n=1...N_e$, as solution of the matter part in \eqref{Hfull}, 
\begin{equation}
 H^{(D)}_{xq,e}= -\frac{\hbar^2}{2m_e}\frac{\partial^2}{\partial x^2} + V(x).
\label{H_e}
\end{equation}
In our electron device scenario, understood as an open system, the Landauer scattering states are such eigenstates. In a closed system one could consider that only ground and first excited states ($n=0,1$) could be relevant for an initial Rabi-like model. However, the number of electron eigenstates here is unlimited: an electron wavepacket, whose energy matches that of the first excited electronic state of the RTD, is injected from $x_{left}$ (see Fig. \ref{volume}), then it is affected by a quantized cavity field in resonance with the electronic level splitting, $\omega=\omega_e = (E_1 - E_0)/\hbar$ and, in a strong coupling regime with high enough $\alpha$, a polaritonic signature is expected to be seen, for example, in the transmission probability of the RTD, as we hope our initial model to be able to capture it. On the other hand, the field part in \eqref{Hfull},
\begin{equation}
 H^{(D)}_{xq,\omega} =  - \frac{\hbar \omega}{2} \frac{\partial^2}{\partial q^2 } + \frac{\hbar \omega}{2} q^2,
\label{H_R}
\end{equation}
is that of an harmonic oscillator, whose eigenstates $\phi_m^{(\omega)}(q)$, $m=0...N_{\omega}$, are well-known (as the Fock number basis), with energies $\hbar \omega (m + 1/2)$. 

A general solution of \eqref{Hfull} can then be written as \cite{QEDFT_refereeexp_2018}
\begin{equation}
\psi(x,q,t) = \sum_n^{N_e} \sum_m^{N_{\omega}} c_{n,m}(t) \phi_n^{(e)}(x) \phi_m^{(\omega)}(q),
\label{posWF}
\end{equation}
which can be recast into
\begin{eqnarray}
\psi(x,q,t)  &=& \sum_m^{N_{\omega}} \Phi_m(x,t) \phi_m^{(\omega)}(q), 
\label{posWF2} \\
\Phi_m(x,t) &=& \sum_n^{N_e} c_{n,m}(t) \phi_n^{(e)}(x) .
\label{posWF2_e}
\end{eqnarray}
Using \eqref{posWF2} in \eqref{Hfull} and integrating out the photon degree of freedom (by multiplying by $\phi_n^{(\omega)}(q)$ and using the orthogonality of the harmonic eigenstates), one obtains
\begin{eqnarray}
&&i\hbar\frac{\partial }{\partial t} \Phi_n(x,t) = \left[ H^{(D)}_{xq,e} + \hbar \omega (n + 1/2) \right] \Phi_n(x,t) \nonumber \\
&+& \alpha x \left[ \sqrt{n+1}\Phi_{n+1}(x,t) + \sqrt{n}\Phi_{n-1}(x,t) \right].
\label{posHproj}
\end{eqnarray}
Equation \eqref{posHproj} states how the interaction part in \eqref{Hfull},
\begin{equation}
H^{(D)}_{xq,i} = \sqrt{2} \alpha q x,
\end{equation}
introduces a set of coupled equations for the electron dynamics, each one refering to a specific photon number, and connected to two neighboring states. In situations where one is interested in spontaneous processes in the cavity field, only two ($N_\omega = 1$) consecutive photon states are relevant, say, $l$ and $l+1$ for any $l$, so that \eqref{posWF2} reads
\begin{equation}
\psi(x,q,t)=\Phi_l(x,t)\phi^{(\omega)}_l(q)+\Phi_{l+1}(x,t)\phi^{(\omega)}_{l+1}(q),
\label{posFAFB_state}
\end{equation}
while \eqref{posHproj} yields 
\begin{eqnarray}
i\hbar \frac{\partial }{\partial t} \Phi_l(x,t)    &=& \left[ H^{(D)}_{xq,e} + \hbar \omega (l+\frac{1}{2}) \right] \Phi_l(x,t)      \nonumber\\[0.5pt]   &+& \sqrt{l+1} \alpha x \Phi_{l+1}(x,t), \nonumber \\
i\hbar \frac{\partial }{\partial t} \Phi_{l+1}(x,t) &=& \left[ H^{(D)}_{xq,e} + \hbar \omega (l+\frac{3}{2}) \right] \Phi_{l+1}(x,t) \nonumber\\[0.5pt] &+&  \sqrt{l+1} \alpha x \Phi_l(x,t),    
\label{posFAFB}
\end{eqnarray}
which describe processes involving spontaneous absorption ($l\to l+1$) and emission ($l+1\to l$) of a photon.

As for the semiclassical analysis the working equation dictated by \eqref{HSC2} becomes, by both neglecting $\epsilon_{dip}$ and including a factor $\sqrt{N}$ as done when reaching \eqref{Hfull},
\begin{equation}
H^{(SC)}_{xq}(t) = -\frac{\hbar^2}{2m_e} \frac{\partial^2}{\partial x^2} + V(x) + \sqrt{2} \alpha q(t) x,
\label{fullSC}
\end{equation}
since the degree of freedom $q$ becomes a time-dependent variable $q(t)$. Differently from \eqref{Hfull}, it yields a $1$D equation only for the electron. For simplicity we ignore the self-consistence between transport and classical Maxwell equations, that is, light affects the electrons, but the electrons do not react to light in this semiclassical picture.

\section{Results}
\label{res}

In this section we show numerical results for the total (particle plus displacement) current at THz frequencies in a RTD inside an optical cavity in the strong coupling regime. Even without light-matter interaction we notice that, with few exceptions, little attention has been devoted to the displacement current in RTDs. In the sequential tunnelling regime, without quantization of electromagnetic fields, predictions on RTD displacement currents, even at frequencies larger than the inverse of the dwell time, have been presented and validated with experimental results \cite{Fei00, Fei00a, Fei01,Fei07,Fei11a, Fei12,Fei11b}. These works have already shown a dramatic dependence of the (complex) admittance of the RTD on the displacement current. Our analysis will also employ such small-signal admittance of the RTD inside a microcavity in Sec. \ref{results_D_AC} for an AC scenario but, first, we will focus on DC behavior in Secs. \ref{results_numerical_transport} and \ref{results_D_DC} to verify whether or not our qualitative model captures polaritonic signatures on transport features.

\subsection{Numerical data}
\label{num}

\begin{figure}
\includegraphics[scale=0.25]{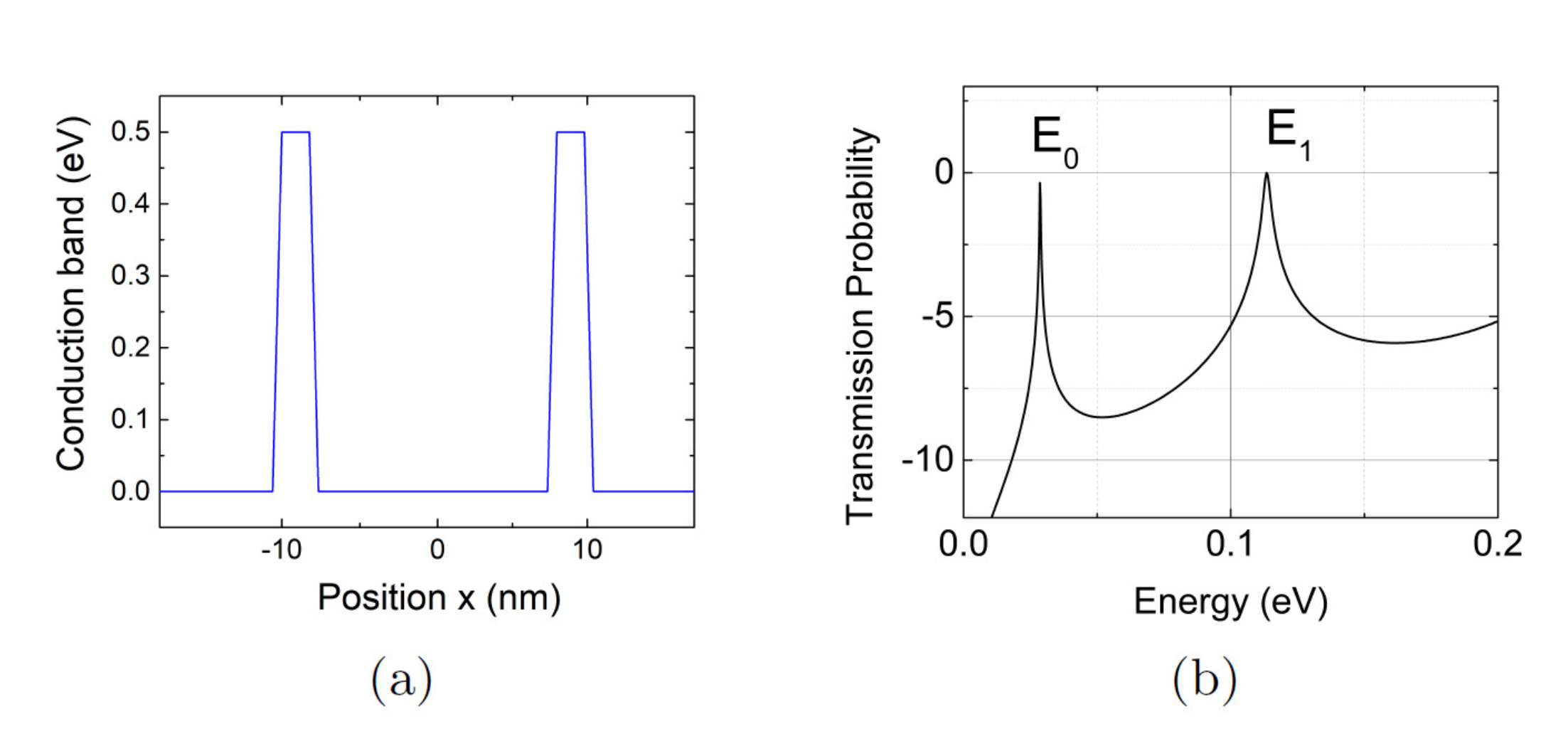}
\caption{(a) Potential profile of the simulated RTD device as function of position $x$; well (barriers) width is $16$ nm ($2$ nm). (b) Transmission probability $T(E)$ from \eqref{T}; only ground and first excited electron states are shown at energies $E_0=28$ meV and $E_1=106$ meV.}
\label{Ec_fig}
\end{figure}

To handle the proposed geometry of Fig. \ref{volume} we consider in the transport direction the symmetric double barrier static potential $V(x)$ of Fig. \ref{Ec_fig}(a), with $L_x = 16$ nm and $L_y,L_z \gg L_x$ assumed to be taken care in advance. Later on we will include a time-dependent THz external signal to such $V(x)$. The barrier width is $L_B=2$ nm and the barrier height is $500$ meV, which are typical parameters of a semiconductor RTD with InGaAs for the well and AlAs for the barriers; for simplicity we consider constant effective mass $m_e=0.041$ $m_0$, with $m_0$ the free electron mass. The transmission coefficient $T(E)$ in Fig. \ref{Ec_fig}(b) shows two resonant energies, $E_0= \hbar \omega_0=28$ meV and $E_1= \hbar \omega_1=106$ meV, whose separation $\hbar \omega_e = \hbar(\omega_1-\omega_0)=78$ meV will then imply a resonant single mode oscillation frequency $\omega = \omega_e \approx 118$ Trad/s ($\nu = \omega/(2\pi)\approx 19$ THz). As anticipated in Fig. \ref{volume} a cavity with sizes around tens of $\mu$m is then required, so that the wavelength approximation also proves valid in our work. We remind that the light-matter coupling constant is turned on only inside the active region, as to also fulfill the requirement of simultaneous conservation of energy and momentum, which drastically minimizes the probabilities of electron-photon interband/intraband collisons in the bulk. As discussed in Sec. \ref{WB}, such coupling constant can be properly engineered so that the RTD works in the strong coupling regime, and for that we assume $\alpha = 1.33$ meV/nm. From the discussion in \eqref{Hfull} one can then estimate the Rabi splitting $E_r = \hbar \omega_r = \alpha L_x \approx 21$ meV, which yields the Rabi frequency $\omega_r = E_r / \hbar \approx 32$ Trad/s ($\nu_r = \omega_r/(2\pi)\approx 5$ THz). These estimations yield in our work $\zeta \equiv \omega_r /\omega \approx 1/4$, confirming that our system is in the strong coupling regime. We remind tough that we are not concerned with exact values for $N$, $L_c$, as in our qualitative model $\alpha$ becomes an input parameter.

\subsection{Wave packet evolution for DC scenarios}
\label{results_numerical_transport}

As mentioned in the Introduction, the effect of electromagnetic fields on the electron current in such devices has been studied through simple collision/scattering models, which account for spontaneous emission/absorption processes, but which are valid only at the very weak coupling regime $\zeta \ll 0.1$ since they neglect the coherence between electron and electromagnetic fields. In this work, on the contrary, in which $\zeta \approx 1/4$ we want exactly to explore the strong coupling regime through a simpliflied model, which however should be used with care at ultrastrong or deep ultrastrong regimes. In Appendices \ref{semic} and \ref{quant} we compare our results respectively to typical Rabi and Jaynes-Cummings models for closed systems, which commonly handle the weak coupling regime $\zeta \le 0.1$.

We consider the case $l=0$ in \eqref{posFAFB_state},
\begin{equation}
\psi(x,q,t) = \Phi_0(x,t) \phi_0^{(\omega)}(q)+\Phi_1(x,t) \phi_1^{(\omega)}(q),
\label{posWF2_results}
\end{equation}
and in \eqref{posFAFB},
\begin{eqnarray}
i\hbar \frac{\partial }{\partial t} \Phi_0(x,t)   &=& \left[ H^{(D)}_{xq,e} + \frac{\hbar \omega}{2} \right] \Phi_0(x,t)         + \alpha x \Phi_{1}(x,t), \nonumber \\
i\hbar \frac{\partial }{\partial t} \Phi_{1}(x,t) &=& \left[ H^{(D)}_{xq,e} +  \frac{3\hbar \omega}{2} \right] \Phi_{1}(x,t) + \alpha x \Phi_0(x,t),   
\label{model}
\end{eqnarray}
where only ground $\phi_0^{(\omega)}(q)$ (zero photon) and first excited $\phi_1^{(\omega)}(q)$ (one photon) states play a role in the transverse fields. The corresponding states $\Phi_0(x,0)$ and $\Phi_1(x,0)$ are initial Gaussian electron wave packets deep inside the reservoirs (with no presence probability in the active region) with spatial dispersion $\sigma_x=150$ nm and initial central position $x_{left}=3.5 \; \sigma_x$ from the active region (Fig. \ref{volume}(c)). The spatial uncertainty corresponds to standard deviation of the energy distribution (in the initial flat potential) of around $5$ meV, such that each initial electron wave packet has a well-defined $E$. The time-evolution done inside a simulation box of $2$ $\mu$m, but outside the active region of about $20$ nm, has to be understood as a mathematical technique to get correct boundary conditions at the borders of the RTD, avoiding spurious reflections when electrons leave it and providing a reasonable description on what type of wave function enters it. As mentioned above, only inside the active region we include light-matter coupling through $H^{(D)}_{xq}$ in \eqref{Hfull}; outside, electrons evolve following $H_{xq,e}^{(D)}$ in \eqref{H_e} decoupled from the electromagnetic fields which follow $H_{xq,\omega}^{(D)}$ in \eqref{H_R}. 

From the quantum model in \eqref{posWF2_results}-\eqref{model} we evolve $\psi(x,q,t)$ and compute the electron probability $P_{Q}(x,t)$ as function of time and position by integrating the degree of freedom $q$, 
\begin{equation}
P_{Q}(x,t)=\int dq |\psi(x,q,t)|^2=|\Phi_0(x,t)|^2+|\Phi_1(x,t)|^2.
\end{equation}
We also evolve a single electron wave packet $\psi(x,t)$ from the semiclassical model $H^{(SC)}_{xq}(t)$ in \eqref{fullSC}, with $q(t) = \text{sin}(\omega t)/\sqrt{2}$, and compute the probability 
\begin{equation}
P_{S}(x,t)=|\psi(x,t)|^2,
\end{equation}
keeping in mind that such sinusoidal oscillation is related to a resonant \textit{internal} electromagnetic field, different from the external signal to be later on applied to $V(x)$. For the sake of comparison we also calculate the probability without light-matter interaction, $P_{0}(x,t)=|\psi(x,t)|^2$, obtained by evolving $\psi(x,t)$ with only $H_{xq,e}^{(D)}$ in \eqref{H_e}. In these three scenarios, the initial energy of the wave packet is that of the first excited electron state, $E=E_1=106$ meV. And, for the quantum case, we assume that initially only $\Phi_{0}(x,0)$, linked to zero photon, is present. 

\begin{figure}	
\includegraphics[scale=0.35]{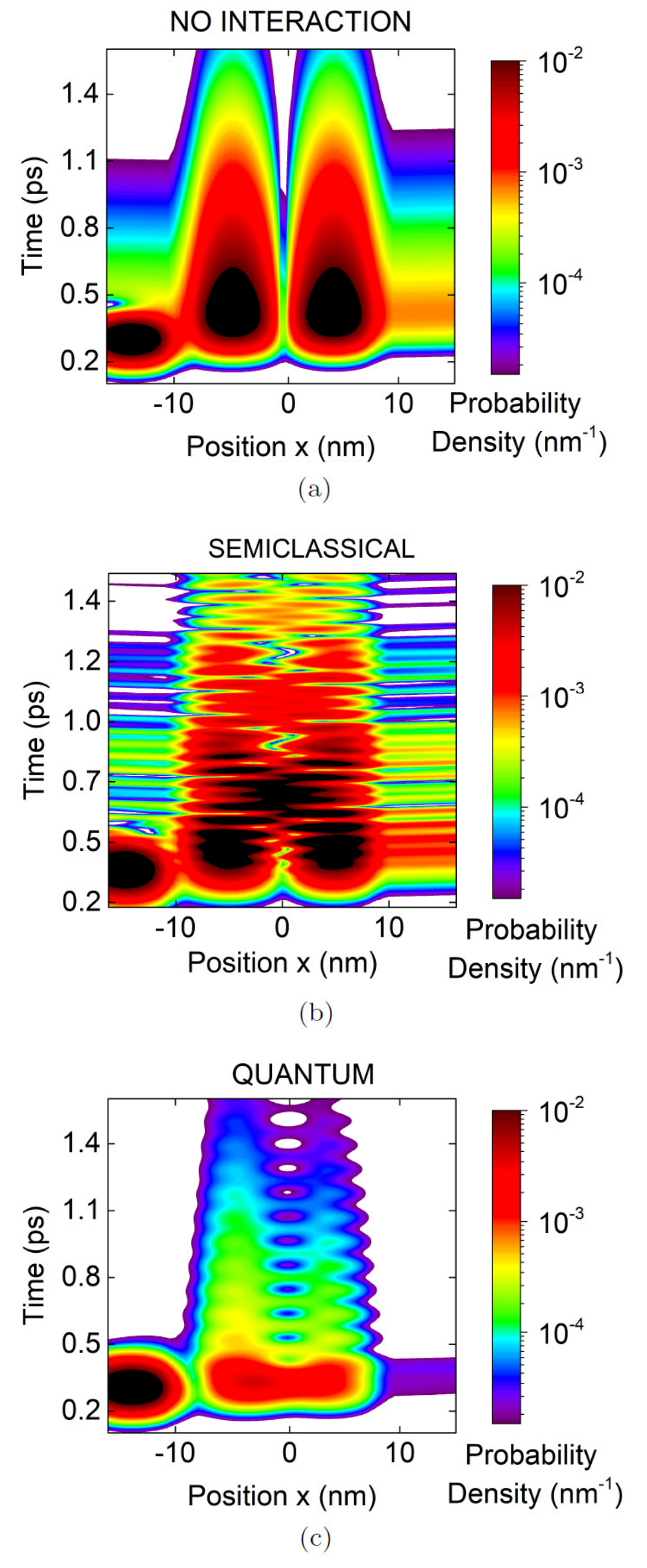}
\caption{(a) Evolution of electron probabilities for three scenarios. $P_{0}(x,t)$ for no light-matter interaction $H_{xq,e}^{(D)}$ in \eqref{H_e}. (b) $P_{S}(x,t)$ for semiclassical interaction $H_{xq}^{(SC)}(t)$ in \eqref{fullSC}. (c) $P_{Q}(x,t)$ for quantum interaction $H_{xq}^{(D)}$ in \eqref{posWF2_results}-\eqref{model}.}
\label{Numerical_quant_probs_alpha25_lev2}
\end{figure}	

In Fig. \ref{Numerical_quant_probs_alpha25_lev2}(a) $P_{0}(x,t)$ shows that, without light-matter interaction, only the first excited state remains occupied inside the RTD, and the wavefunction gets mostly transmitted. The semiclassical case $P_S(x,t)$ in Fig. \ref{Numerical_quant_probs_alpha25_lev2}(b) shows two different frequencies: first, there is a fast oscillation with frequency $\nu$ due to the resonant term $\approx \text{sin}(2 \pi \nu t)$, which is adiabatically activated when the wave packet enters the well; second, there is a slower natural oscillation at the Rabi frequency $\nu_r$.

On the other hand, for the quantum case $P_{Q}(x,t)$ in Fig. \ref{Numerical_quant_probs_alpha25_lev2}(c), only the Rabi oscillation seems to play a role by periodically transitioning between two RTD states but with a frequency that is about $2 \nu_r$, while the fast resonant oscillations at frequency $\nu$ seem to be missing. Such features seem to expose the \textit{polaritonic} nature of the process: thanks to the cavity field, electrons no longer evolve as superposition of the pure electronic eigenstates $|0\rangle,|1\rangle$ of $H_{xq,e}^{(D)}$ but, instead, of the light-matter eigenstates $|\psi_{\pm} \rangle=(|1,0\rangle \pm |0,1\rangle)/\sqrt{2}$ of $H^{(D)}_{xq}$. Figure \ref{volume}(d) depicts that the decoupled $|1,0\rangle$ and $|0,1\rangle$ states, degenerate at zero coupling (dashed line), acquire a $E_+-E_-=2 E_r$ splitting at nonzero coupling; in Appendix \ref{dressed} we show the typical closed-system diagonalization of such polaritonic subspaces and address such avoided crossings.

\subsection{Transmission coefficient $T(E)$ for DC scenarios}
\label{results_D_DC}

We now address the transmission coefficient $T(E)$ from \eqref{T} to verify wheter or not the polaritonic feature can really be seen in such transport feature. Following the same scenarios as described by $P_{0,S,Q}(x,t)$, we compare the three scenarios for $T(E)$, that is, $T_0(E)$, $T_S(E)$, $T_Q(E)$, which refer respectively to no light-matter, semiclassical, quantum interaction scenarios.

\begin{figure}	
\includegraphics[scale=0.35]{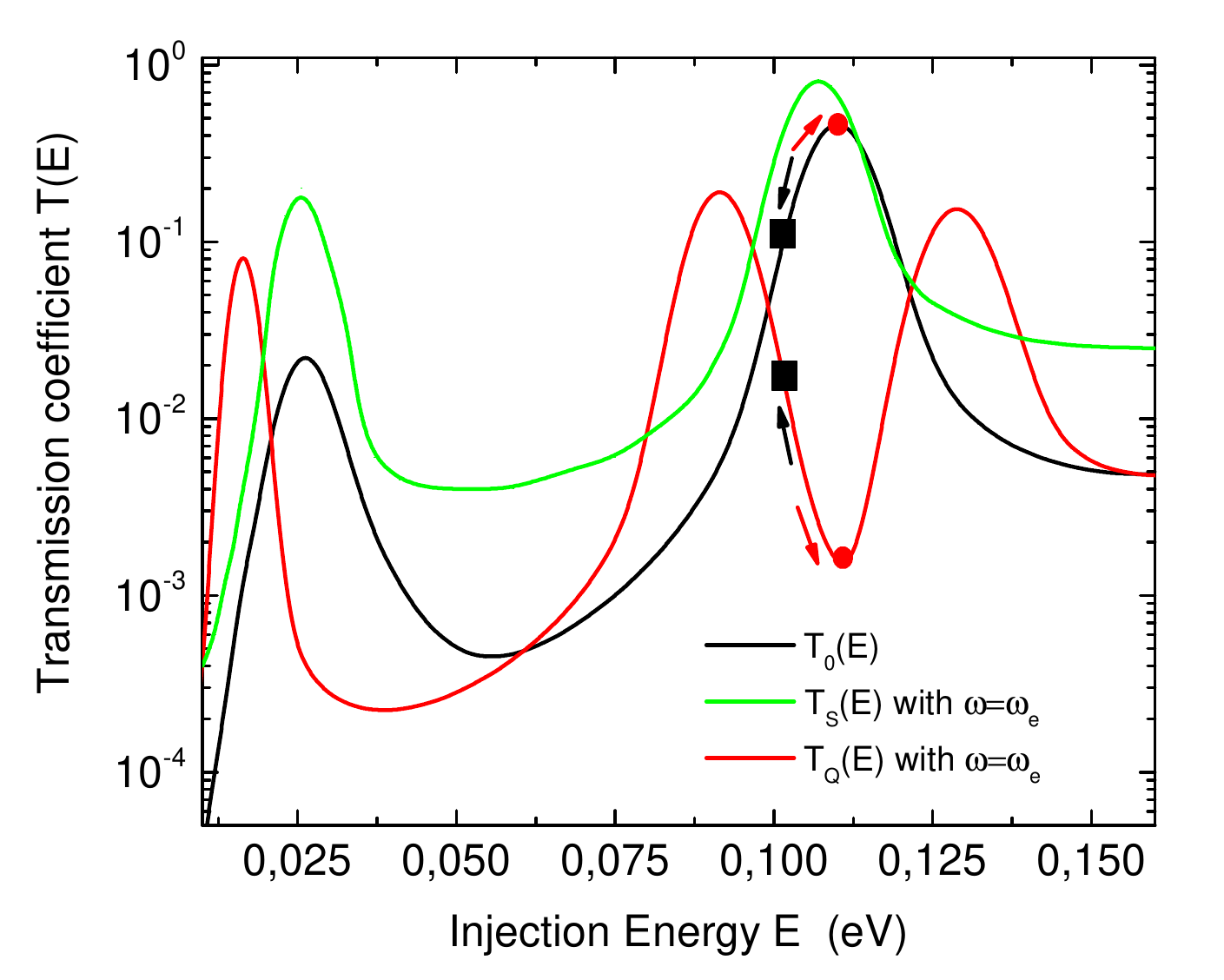}
\caption{Transmission coefficients as function of energy for three different scenarios: (black) no light-matter interaction $T_0(E)$, (green) semiclassical interaction $T_S(E)$ for a resonant photon $\omega=\omega_e$, and (red) quantum interaction $T_Q(E)$ for a resonant photon $\omega=\omega_e$. The symbols and arrows will be used to explain Fig. \ref{D_in_f}.}
\label{D(E)_vs_T(E)}
\end{figure}

In Fig. \ref{D(E)_vs_T(E)} we compare $T_0(E)$ (black) with $T_S(E)$ (green, resonant $\omega=\omega_e$). No important differences are seen, apart from a higher transmission coefficient for $T_S(E)$. This small difference could be explained by knowing that in the semiclassical picture the bottom of the well is oscillating at $\nu\approx19$ THz, so that electrons with an initial low energy are, at some times, aligned with the resonant energies of the well, so providing more transmission than with $\nu=0$. One could be tempted to say that this reflects recent works showing that strong light-matter coupling may enhance conductivity \cite{conductivity2DEG,conductivity2DEGnon,conductivitylandau, conductivityorgan}, but one should also keep in mind the quantitative limitations of our model.

However, $T_Q(E)$ (red, resonant $\omega=\omega_e$) in Fig. \ref{D(E)_vs_T(E)} presents as striking new feature the appearance of two peaks around the energy of the first excited state of the RTD. As previously discussed, this is the signature of the \textit{avoided} crossing $E_+ - E_- = 2 \hbar  \omega_r\approx 42$ meV. The true splitting in Fig. \ref{D(E)_vs_T(E)} is slightly different likely due to the broadening of the states in such open system. This is the first main result of our work: although not new in other platforms, and somehow expected, transport properties may indeed indicate the influence of a cavity quantized field over the electrons crossing an RTD in the strong coupling regime, so revealing their polaritonic nature.

No such polaritonic splitting appears around the energy of the ground state of the RTD since we considered $l=0$, and Fig. \ref{volume}(d) and \eqref{posHproj} indicate that the $|0,0\rangle$ state remains uncoupled, while \textit{every} other state can sustain an avoided crossing: had we included more states we would also have a $|1,1\rangle$-$|0,2\rangle$ coupling, and so on. Not including more polariton states in the RTD, and limiting the coupling to only among two consecutive states when going from \eqref{posHproj} to \eqref{posFAFB}, are simplifications that limit the use of our model into even stronger coupling regimes or for treating nonresonant cases at large frequencies.

\subsection{Displacement current coefficient $D^{(f)}(E,t)$ for AC scenarios}
\label{results_D_AC}

We now include in our model an external AC small-signal battery $V_{in}(t)=V_{A} \text{cos}(2\pi  f t+ \Theta)$ between emitter and collector, in addition to the previous DC bias. We assume the emitter to be grounded so that the scalar potential $V(x,t)$ inside the RTD (the bottom of the conduction band) becomes time-dependent,
\begin{equation}
V(x,t)=V(x)+ V_{A} \; \text{cos}(2\pi \; f \; t+ \Theta) \; \frac{x+L/2}{L},
\label{Ec}
\end{equation}
with $V(x)$ the same as in Fig. \ref{Ec_fig}(a); the range of $f$-values covers $0.2 \le f \le 2$ THz, the voltage amplitude $V_A=10$ mV ensures the small-signal regime, and the role of the phase $\Theta$ will be later explained. We remind that $V_{in}(t)$ has no relation with the sinusoidal oscillation present in the semiclassical model (whose resonant frequency is $\nu \approx 19$ THz). The term $(x+L/2) / L$ in \eqref{Ec} is an ad-hoc (not self-consistent) interpolation of a reasonable potential profile. 

To understand how we will represent $D^{(f)}(E,t)$, we briefly introduce the small-signal (complex) admittance $Y_{11}(f)$, defined as
\begin{eqnarray}
Y_{11}(f)=\frac{I_0^{(f)}(0)-I_{DC}}{V_A} + i \frac{I_{-\pi/2}^{(f)}(0)-I_{DC}}{V_A},
 \label{Y11s}
\end{eqnarray}
where $I_{0}^{(f)}(0)$ and $I_{-\pi/2}^{(f)}(0)$ are the total current when the external voltage is respectively $V_A  \text{cos} (2\pi  f  t)$ and $V_A  \text{cos} (2\pi  f  t-\pi/2)$. The computation of $I^{(f)}(0)$ comes from \eqref{curfinal2} and \eqref{I_Fermi}. As already discussed, the second and third terms in \eqref{equation_7} can be neglected in the $f$-range of our simulation, which is no more than $10 \%$ of the relevant (resonant) frequencies of the vector potential. As mentioned in Sec. \ref{dis}, the total current measured at time $t=0$ is due to a flux of electrons that have entered in the active region during the time interval $[-\tau,0]$, being $\tau$ the average electron transit time; thus, the last electron entering the active region at time $t=0$ has \textit{seen} only the external potential $V(x,0)$, while the first electron entering at time $t=0-\tau$ has \textit{memory} of the whole evolution from $V(x,-\tau)$ until $V(x,0)$.

 In principle, the computation of $I^{(f)}(0)$ in \eqref{I_Fermi} requires also an energy integral of $D^{(f)}(E,0)$ but, since we are interested on the dynamics around $E=E_1$, the energy integral is ignored. As such (apart from some irrelevant constants) we can identify $I^{(f)}(0) \to D^{(f)}(E,0)$ and $I_{DC} \to T(E)$, so that \eqref{Y11s} is rewritten here as
\begin{eqnarray}
Y_{11}(f)=\frac{D^{(f)}_0(E,0)-T(E)}{V_A} + i \frac{D^{(f)}_{-\pi/2}(E,0)-T(E)}{V_A}.
\label{Y11}
\end{eqnarray}
So we once more consider the electron injection energy at $E=E_1=106$ meV and take the initial state to be $\psi(x,q,0)=\Phi_0(x,0)\phi_0^{(\omega)}(q)$. Since the coupled equations in \eqref{model} are inherently time-dependent, computing the effects of $V(x,t)$ on $\psi(x,q,t)$ is straigthforward: once we get  $\psi(x,q,t)$ for each relevant injection time $t_i$, from its current density, we compute $D^{(f)}(E,0)$ by using \eqref{D_definition_60}. 

\begin{figure}
\includegraphics[scale=0.7]{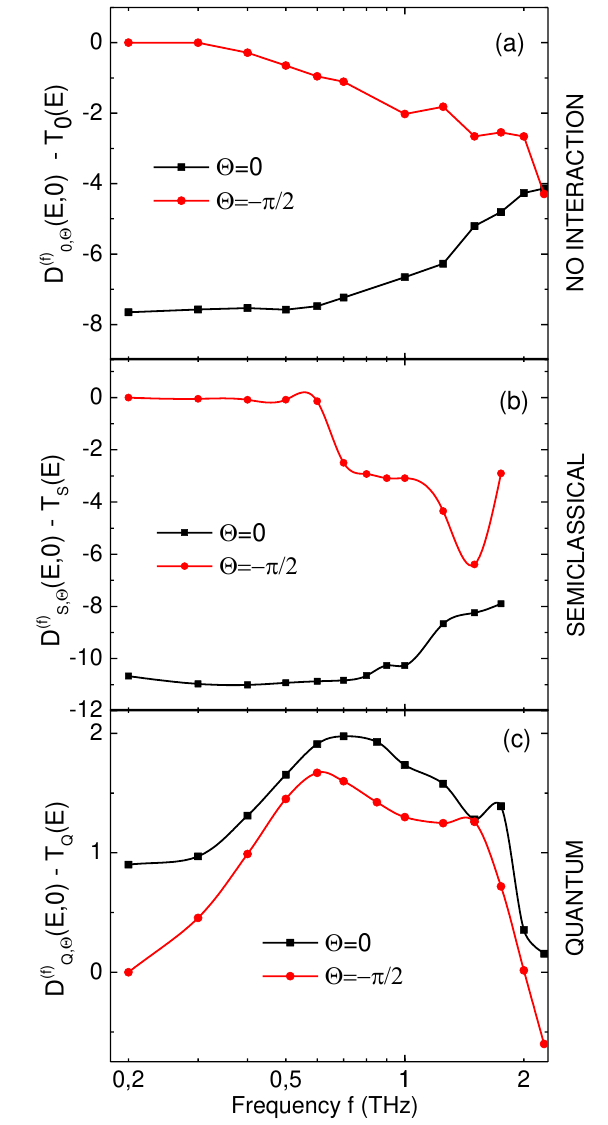}
\caption{Dependence of $D^{(f)}(E,0)-T(E)$ on external frequency $f$ of an AC small signal ($V_A=10$ mV) in three scenarios: (a) no light-matter interaction $D_{0,\Theta}^{(f)}(E,0)$; (b) semiclassical interaction $D_{S,\Theta}^{(f)}$(E,0); (c) quantum interaction $D_{Q,\Theta}^{(f)}$(E,0). Electrons are injected with the DC resonant energy $E=E_1=106$ meV. Two different phases of the input signal, $\Theta=0$ (black) and $\Theta=-\pi/2$ (red), are considered.}
\label{D_in_f}
\end{figure}

However, the interpretation of $D^{(f)}(E,0)$ is more convoluted than $T(E)$ because, strictly speaking, dealing with an explicit time-dependent $V(x,t)$ means that neither electron nor polariton well-defined resonant energies exist anymore. A reasonable way to infer the memory effects on $D^{(f)}(E,0)$ is to evaluate $T(E)$ (still with the resonant energies) with the following energy offset,
\begin{eqnarray}
D^{(f)}(E,0) \propto T\left(E-\frac{V_{A} \; \text{cos}(-2\pi \; f \; \tau+ \Theta)|e|}{2}\right) , 
\label{offset}
\end{eqnarray}
where $-\tau$ correspond to the injection time of the \textit{last electron}. Such offset is the result of assuming that the net effect of the spatial potential in $V(x,t)$ in \eqref{Ec} can be assimilated by moving the bottom energy in the barrier region of $V(x)$ by half of the AC bias. Of course, the assumption in \eqref{offset} is only reasonable when $f \; \tau< 1$. With the tool in \eqref{offset} we can analyze the results in Fig. \ref{D_in_f} for the three scenarios discussed so far, that is, no light-matter, semiclassical, and quantum interactions, which we respectively label as $D_{0,\Theta}^{(f)}$, $D_{S,\Theta}^{(f)}$, and $D_{Q,\Theta}^{(f)}$; the two phases $\Theta=0$ and $\Theta=-\pi/2$ mentioned above are considered in each case. An immediate feature in the plots of $D^{(f)}(E,0)$ in Fig. \ref{D_in_f} is that, similar to the plots of $T(E)$ in Fig. \ref{D(E)_vs_T(E)}, the semiclassical and no light-matter interaction results have similar tendencies, while the behavior of the quantum case is very different. 

For low frequencies, $f \tau \approx  0$,  the value $D_{-\pi/2}^{(f)}(E_1,0)$ (red curve) in the three panels starts at zero because $\text{cos}(-\pi/2)=0$ in \eqref{offset} so that $D_{-\pi/2}^{(f)}(E_1,0)\propto T(E_1)$. In other words, the energy of the red circle in Fig. \ref{D(E)_vs_T(E)} without the small signal corresponds to the energy $E_1$ of electrons contributing to $D_{-\pi/2}^{(f)}(E_1,t)$ with the small signal. 

On the other hand, the value $D_{0}^{(f)}(E_1,0)$ (black curve) is higher in the non-interaction and semi-classical cases, while lower in the quantum case, because $\text{cos}(0)=1$ in \eqref{offset} when $f \tau \approx  0$ and $\Theta=0$. This means that the energy $E_1-V_A|e|/2$ of the black square in Fig. \ref{D(E)_vs_T(E)} without small signal corresponds to the energy $E_1$ of electrons contributing to $D_{-\pi/2}^{(f)}(E_1,t)$ when the small signal, with $\Theta=0$, is included. The net result is that  $D_{0,S}^{(f)}(E_1,t) \propto T_{0,S}(E_1-V_A|e|/2)<T_{0,S}(E_1)$ in the non-interaction and semi-classical cases, while $D_{Q}^{(f)}(E_1,t) \propto T_Q(E_1-V_A|e|/2)>T_Q(E_1)$ in the quantum case. All these results just tell us that, at low frequency, the complex small-signal admittance $Y_{11}(f)$ is, in fact, a real number (conductance). At such frequencies, $f \tau \approx  0$, and this well-known result can be understood by noticing that there is no big difference between $V(x,0)$ and $V(x,-\tau)$, because electrons evolve much faster than the rate of change of the external signal. There is no \textit{memory} effect from the displacement current in the total current. 

Such \textit{memory} effects start to be present when $\tau \ge 1/f$ so that $Y_{11}(f)$ has non-zero real and imaginary parts. In such conditions, we see that the red curve starts to deviate from the low frequency results faster than the black curve. For $\tau \ge 1/f$, the evolution of $V(x,t)$ in the time interval $[-\tau,0]$ is relevant for the value of the current measured at time $t=0$. When $\Delta \Theta=-2\pi f\tau $ is small, but not zero, a Taylor expansion of the small signal voltage in \eqref{offset} gives $\text{cos}(\Delta \Theta-\pi/2)=\text{sin}(\Delta \Theta)\approx\Delta \Theta$ (red curve) and $\text{cos}(\Delta \Theta)\approx1-\Delta \Theta^2$ (black curve). In other words, the slope of $\text{sin}(\Delta \Theta)$ (red curve) is greater than the slope of $\text{cos}(\Delta \Theta)$ (black curve) as seen in Fig. \ref{D_in_f} because $|\Delta \Theta|>|\Delta \Theta^2|$. The arrows in Fig. \ref{D(E)_vs_T(E)} try to indicate which are the energies of the DC transmission coefficient (still the resonant energies) that are relevant for the computation of the displacement current coefficient in Fig. \ref{D_in_f}. The arrows start at an energy in \eqref{offset} corresponding to $\text{cos}(-2\pi f\tau+\Theta)$ for the \textit{last electron} injection time $-\tau$, while it always end up at an energy corresponding to $\text{cos}(\Theta)$ for the \textit{first electron} injection time $0$. We notice that for higher frequencies the computational scenarios are quite complex, so that the simple Taylor explanations from \eqref{offset} are no longer useful. 

Two are the main conclusions from Fig. \ref{D_in_f}. First, we clearly see the polaritonic nature of the light-matter interaction in the complex small-signal admittance $Y_{11}(f)$. The small-signal procedure has empirical advantage, when compared to the direct computation of the total current $I^{(f)}(t)$, as we can expect that evaluating $Y_{11}(f)$ around the energy $E_1$ is reasonable as far as the device is properly polarized and $V_A$ is small enough. On the other hand, the polaritonic nature of $T(E)$ seen in Fig. \ref{D(E)_vs_T(E)} around $E_1$ could somehow be masked in the whole energy integral required in \eqref{I_Fermi}. Second, the high frequency behavior of $Y_{11}(f)$ shows that there is plenty of room for engineering new THz applications in electronic or optoelectronic devices when the displacement current with coherent electron-photon interaction is taken into account.  

\section{Conclusions}
\label{con}

When one is interested in getting information on the dynamic behavior of electron devices at frequencies around the inverse of the electron transit time in the active region, electrostatic approximations are no longer valid since at such high frequencies what is measured is the total current, the sum of particle plus displacement components. One is forced to consider time-dependent scenarios and include the role of transverse electromagnetic fields present in the active region. In many scenarios, a self-consistent coupling between quantum electron transport models and classical electromagnetic fields is enough to account for the main physics. In other scenarios, effects of a quantized electromagnetic field on electron dynamics can be translated as collisions (absorbing or emitting photons) through Fermi Golden Rule. However, in scenarios where a resonant strong light-matter coupling regime is induced, perturbative approaches are no longer applicable and a quantum treatment of both electrons and electromagnetic fields is needed. 

In this work, by focusing on such strong coupling regime, transport properties of RTD devices operating at the THz gap placed inside a resonant cavity have been studied. We have shown that the quantization of the cavity electromagnetic field is needed, in addition to the quantization of the matter, because the entangled nature of the light-dressed states of matter, the so-called polaritons, can in principle be experimentally probed in transport measurements. The results from our electron transmission coefficients show that a full quantum treatment may indicate the existence of polaritons once electrons are crossing a resonant cavity, a feature absent in the semiclassical picture where the field is taken with no backaction from the matter. We have also developed a new AC displacement current coefficient to substitute the usual DC transmission coefficient. When applying an external signal to the RTD, this new coefficient also reveals the polaritonic nature of the system at high frequency.

The Bohmian description \cite{Bohm1,Bohm2,Xavier_book,Xavier_Ferry} employed in this work, for the quantum case, assumes that positions of electrons and amplitudes of electromagnetic fields inside the active region are well-defined. Such assumptions are not rigorous within the Orthodox description because such knowledge is only real, strictly speaking, when a measuring apparatus is explicitly considered. Our use of a pure wave function is not to be understood as a naive way of studying an open quantum system, which requires the use of the reduced density matrix in the Orthodox description, because the concept of Bohmian conditional wave function \cite{Xavier_PRL,Norsen} shows that such pure states are a perfectly valid conceptual tool. 

Some approximations had to be done so that our model, initially developed for many-electrons and many-modes, could yield a simple initial picture of such a complex problem, like considering a single mode for the field, an effective single-electron picture for electrons in the conduction band in the effective mass approximation, a two-level picture for the electrons although not based on eigenstates but rather in scattering states and, mostly, assuming from the very beginning the long wavelength approximation. From a `similar' infinite quantum well system, Appendix \eqref{appendixres} makes comparisons towards typical closed system models as Rabi and Jaynes-Cummings, under Rotating Wave Approximation. 

Although due to such approximations this simplified version of our model cannot be used without care in even stronger coupling regimes, and although these qualitative results are not exactly new or unexpected in solid state \cite{QWsplit92,RTD_cavityexp}, and well-known in other platforms \cite{vacuum_chem1,vacuum_chem2,vacuum_chem3,vacuum_chem4} or in the quantum optics community \cite{grynberg, scully}, we do believe that they may open original unexplored paths for engineering new electron devices and new applications in the THz gap by taking profit of the interplay between quantum electrons and quantized electromagnetic fields.

\begin{acknowledgments}
Research funded by Spain's Ministerio de Ciencia, Innovaci\'on y Universidades under Grant RTI2018-097876-B-C21 (MCIU/AEI/FEDER, UE), Grant PID2021-127840NB-I00 (MICINN/AEI/FEDER, UE), the ``Generalitat de Catalunya" and FEDER for the project 001-P-001644 (QUANTUMCAT), the European Union's Horizon 2020 research and innovation programme under Grant 881603 GrapheneCore3 and under the Marie Sk\l{}odowska-Curie Grant 765426 TeraApps.
\end{acknowledgments}

\begin{appendix}

\begin{widetext}
\section{Derivation of the effective Bohmian Hamiltonian}
\label{bohmderiv}

This Appendix shows how to connect the full many-electron many-mode Hamiltonian in \eqref{Hdip} with the effective single-electron single-mode Hamiltonian in \eqref{Hfull}, where the latter is obtained after important approximations are implemented. Their connection through the use of Bohmian conditional wave functions shows the way to include some of the many-electron many-mode correlations into \eqref{Hfull} by using Bohmian trajectories explicitly. 

\subsection{Bohmian conditional wave function for light-matter interaction}

We start from the Hamiltonian in Dipole gauge under the long wavelength approximation in \eqref{Hdip}, 
\begin{eqnarray}
H^{(D)}&=&-\frac{\hbar^2}{2m_e}\sum_{j} \mathbf{\nabla}^2_{j} +\frac{1}{4\pi\epsilon_0}\sum_{j< k}\frac{e^2}{|\mathbf{r}_j-\mathbf{r}_k|} +\sum_{j}V(\mathbf{r}_{j})
            +\sum_{n,\lambda}\frac{\hbar \omega_ n}{2} \left(-\frac{\partial^2}{\partial q^2_{n,\lambda}} +q^2_{n,\lambda}\right) \nonumber \\
           &+&\sum_{n,\lambda} \sum_{j}e \sqrt{\frac{\hbar\omega_ n}{\epsilon_0 L_c^3}} q_{n,\lambda} \mathbf{\epsilon}_{n,\lambda}\cdot \mathbf{r}_j
           +\sum_{n,\lambda}  \frac{e^2}{2\epsilon_0 L_c^3} \left(\sum_{j} \mathbf{\epsilon}_{n,\lambda}\cdot \mathbf{r}_j \right)^2,
\end{eqnarray}
where $\mathbf{r}_j=\{x_j,y_j,z_j\}$ is the 3D position of $j=1,..,N$ particles, and $n$ identifies each wave vector of the $n=1,..,M$ electromagnetic modes with $\lambda=1,2$  perpendicular directions. Approximations are needed to tackle such many-body problem. For example, distinct approximations result in distinct `flavours' of the (QE)DFT as cited in our Introduction. We here develop a different approximation, based on the conditional wave function as offered from the Bohmian theory, which interprets the \textit{local-in-position} continuity equation \eqref{conti} as really implying dynamics of \textit{local-in-position} particles.    

\subsubsection{Full system of conditional wave functions coupled by trajectories}

The full $N$-body $M$-mode wave function can be written, by singling out one degree of freedom for each matter and light, as
\begin{equation}
\psi^{x,q}(x_1,y_1,z_1,...,x_N,y_N,z_N,q_{1,1},q_{ 1,2},...,q_{M,1},q_{ M,2},t)=\psi^{x,q}(x,q,\mathbf{w},t),
\end{equation}
where $x\equiv x_1$, $q \equiv q_{ 1,1}$, and $\mathbf{w}\equiv\{y_1,z_1,x_2,y_2,z_2,...,x_N,y_N,z_N,q_{1,2},q_{2,1},q_{2,2},...,q_{ M,1},q_{ M,2}\}$, which is a solution of the Schroedinger equation 
\begin{equation}
i\hbar \frac{\partial}{\partial t} \psi^{x,q}(x,q,\mathbf{w},t)=H^{(D)} \psi^{x,q}(x,q,\mathbf{w},t).
\end{equation}
Following the Bohmian approach, one particular experiment is defined by such wave function $\psi^{x,q}(x,q,\mathbf{w},t)$ and by the ``trajectories'' $x[t]$, $q[t]$, $\mathbf{w}[t]$, where $x[t]$ is really a trajectory of a particle while $q[t]$ is just the evolution in time of the amplitude of the electromagnetic mode. To compute them we need a ``velocity'' that is obtained from the wave function itself. Interestingly, the velocities related to $x[t]$ and $q[t]$ can be computed from either $\psi^{x,q}(x,q,\mathbf{w},t)$ or from the conditional wave function
\begin{equation}
\psi^{x,q}(x,q,t) \equiv \psi^{x,q}(x,q,\mathbf{w}[t],t),
\end{equation}
whose equation of motion obeys
\begin{equation}
\left[ i\hbar \frac{\partial}{\partial t} \psi^{x,q}(x,q,\mathbf{w},t)\right]_{\mathbf{w}=\mathbf{w}[t]}=\left[ H^{(D)} \psi^{x,q}(x,q,\mathbf{w},t)\right]_{\mathbf{w}=\mathbf{w}[t]}.
\label{Zeq}
\end{equation}
The left hand side in \eqref{Zeq} yields
\begin{eqnarray}
i\hbar \frac{\partial}{\partial t} \psi^{x,q}(x,q,t)= \left[ i\hbar \frac{\partial}{\partial t} \psi^{x,q}(x,q,\mathbf{w},t)\right]_{\mathbf{w}=\mathbf{w}[t]}+ i\hbar \sum_{i} \left[ \frac{\partial \psi^{x,q}(x,q,\mathbf{w},t)}{\partial w_i} \right]_{\mathbf{w}=\mathbf{w}[t]} \frac{d w_i[t]}{dt},
\label{cal1}
\end{eqnarray}
where $i$ embodies the remaining $3N+2M-2$ terms in $\mathbf{w}$.

Let us compute the right hand side in \eqref{Zeq} for each of the six terms present in $H^{(D)}$. The kinetic energy yelds 
\begin{eqnarray}
\left[-\frac{\hbar^2}{2m_e}\sum_{j=1}^{N} \mathbf{\nabla}^2_{j} \psi^{x,q}(x,q,\mathbf{w},t)\right]_{\mathbf{w}=\mathbf{w}[t]}=-\frac{\hbar^2}{2m_e}\frac{\partial^2 \psi^{x,q}(x,q,t)}{\partial x^2} -\frac{\hbar^2}{2m_e}\left[\sum_{j=1,x_j\neq x_1}^{N} \left(\frac{\partial^2}{\partial x_j^2}+\frac{\partial^2}{\partial y_j^2}+\frac{\partial^2}{\partial z_j^2} \right) \psi^{x,q}(x,q,\mathbf{w},t)\right]_{\mathbf{w}=\mathbf{w}[t]},
\label{cal2}
\end{eqnarray}
where the first term is the kinetic energy of the single degree of freedom $x$, while the second term is a function of $x$, $q$, $t$ that includes the many-body correlations with the other modes and particles. The Coulomb interaction yields 
\begin{eqnarray}
&&\left[\frac{1}{4\pi\epsilon_0}\sum_{k>j=1}^{N}\frac{e^2 }{|\mathbf{r}_j-\mathbf{r}_k|} \psi^{x,q}(x,q,\mathbf{w},t) \right]_{\mathbf{w}=\mathbf{w}[t]}=\frac{1}{4\pi\epsilon_0}\sum_{k \neq 1}^{N}\frac{e^2}{\sqrt{(x-x_k[t])^2+(y_1[t]-y_k[t])^2+(z_1[t]-z_k[t])^2}}\psi^{x,q}(x,q,t) \nonumber\\
&+&\frac{1}{4\pi\epsilon_0}\sum_{k>j=1, j\neq 1}^{N}\left(\frac{e^2}{\sqrt{(x_j[t]-x_k[t])^2+(y_j[t]-y_k[t])^2+(z_j[t]-z_k[t])^2}} \right) \psi^{x,q}(x,q,t), 
\label{cal3}
\end{eqnarray}
where the first term is a potential that depends on $x$ and $t$, while the second term is a potential that depends on $t$ only and it can be ignored. The external potential yields
\begin{eqnarray}
\left[\sum_{j=1}^{N}V(\mathbf{r}_{j}) \psi^{x,q}(x,q,\mathbf{w},t)\right]_{\mathbf{w}=\mathbf{w}[t]}= V(x,y_1[t],z_1[t])  \psi^{x,q}(x,q,t) +\sum_{j \neq 1}^{N} V(x_j[t],y_j[t],z_j[t]) \psi^{x,q}(x,q,t),
\label{cal4}
\end{eqnarray}
where the first term is a potential that depends on $x$ and $t$, while the second term is a potential that depends on $t$ only and it can be ignored.  The electromagnetic energy yields 
\begin{eqnarray}
\left[\sum_{n,\lambda=1,1}^{M,2}\frac{\hbar \omega_{n}}{2} \left(-\frac{\partial^2}{\partial q^2_{n,\lambda}} +q^2_{n,\lambda}\right) \psi^{x,q}(x,q,\mathbf{w},t)\right]_{\mathbf{w}=\mathbf{w}[t]}&=&\frac{\hbar \omega}{2} \left(-\frac{\partial^2}{\partial q^2} +q^2\right) \psi^{x,q}(x,q,t) \\
&+&\left[ \sum_{{ n},\lambda \neq\ {1},1}^{M,2}\frac{\hbar \omega_{n}}{2} \left(-\frac{\partial^2}{\partial q^2_{n,\lambda}} +q^2_{n,\lambda}\right) \psi^{x,q}(x,q,\mathbf{w},t)\right]_{\mathbf{w}=\mathbf{w}[t]}, \nonumber
\label{cal5}
\end{eqnarray}
where the first term is a single-mode Hamiltonian that depends on $q$, while the second term is a function of $x$, $q$, $t$ that includes the many-body correlations with the other modes and particles. The electron-photon interaction term depends on the specific types of electromagnetic modes involved. To simplify the discussion in the development in this appendix, we assume a linear polarization of electric field in $\hat x$ direction and only one $\lambda$, $\mathbf{\epsilon}_{n,\lambda}\cdot \mathbf{r}_j=x_j$, so that     
\begin{eqnarray}
\left[\sum_{n,\lambda=1,1}^{M,2} \sum_{j=1}^{N}e \sqrt{\frac{\hbar\omega_{n}}{\epsilon_0 L_c^3}} q_{n,\lambda} \mathbf{\epsilon}_{n,\lambda}\cdot \mathbf{r}_j \psi^{x,q}(x,q,\mathbf{w},t)\right]_{\mathbf{w}=\mathbf{w}[t]}= e \sqrt{\frac{\hbar\omega}{\epsilon_0 L_c^3}} q x  \psi^{x,q}(x,q,t) \nonumber\\
+ \sum_{j \neq 1}^{N}e \sqrt{\frac{\hbar\omega}{\epsilon_0 L_c^3}} q x_j[t] \psi^{x,q}(x,q,t)+\sum_{{ n} \neq 1}^{M}e \sqrt{\frac{\hbar\omega_{ n}}{\epsilon_0 L_c^3}} q_{n}[t] x \psi^{x,q}(x,q,t)+ \sum_{{n}\neq 1}^{M} \sum_{j \neq 1}^{N}e \sqrt{\frac{\hbar\omega_{n}}{\epsilon_0 L_c^3}} q_{n}[t]x_j[t] \psi^{x,q}(x,q,t),
\label{cal6}
\end{eqnarray}
where the first term is a potential that depends on $q$ and $x$, the second and third terms are potentials that respectively only depends on $q$ or $x$, and the fourth term is another potential that depends on $t$ only and it can be ignored. At last, the dipole self-energy yields  
\begin{eqnarray}
\left[\sum_{n,\lambda=1,1}^{M,2}  \frac{e^2}{2\epsilon_0 L_c^3} \left(\sum_{j} \mathbf{\epsilon}_{n, \lambda}\cdot \mathbf{r}_j \right)^2 \psi^{x,q}(x,q,\mathbf{w},t)\right]_{\mathbf{w}=\mathbf{w}[t]}=  M \frac{e^2}{2\epsilon_0 L_c^3} \left( x+\sum_{j \neq 1}^{N} x_j[t]  \right)^2 \psi^{x,q}(x,q,t),
\label{cal7}
\end{eqnarray}
which depends on $x$ and $t$ and can be considered as a type of potential profile for the electrons. 	

By collecting all potential terms in \eqref{cal1}-\eqref{cal7} we get an effective Schroedinger equation for $x$-$q$ as
\begin{eqnarray}
i\hbar \frac{\partial}{\partial t} \psi^{x,q}(x,q,t)=H^{(D)}_{xq} \psi^{x,q}(x,q,t),
\label{final0}
\end{eqnarray}
where the new Dipole gauge Hamiltonian reads
\begin{eqnarray}
H^{(D)}_{xq}=-\frac{\hbar^2}{2m_e}\frac{\partial^2}{\partial x^2}+ W(x,t)+\frac{\hbar \omega}{2} \left(-\frac{\partial^2}{\partial q^2} +q^2\right)+e \sqrt{\frac{\hbar\omega}{\epsilon_0 L_c^3}} q \left(x + \sum_{j \neq 1}^{N}x_j[t]\right) +G(x,q,t)+i J(x,q,t),
\label{final}
\end{eqnarray}
where $W(x,t)$ includes all potentials that depend on $x$ and $t$,
\begin{eqnarray}
W(x,t) &=&\frac{1}{4\pi\epsilon_0}\sum_{k \neq 1}^{N}\frac{e^2}{\sqrt{(x-x_k[t])^2+(y_1[t]-y_k[t])^2+(z_1[t]-z_k[t])^2}} + V(x,y_1[t],z_1[t])  \nonumber \\
           &+&\sum_{{n} \neq 1}^{M}e \sqrt{\frac{\hbar\omega_{ n}}{\epsilon_0 L_c^3}} q_{ n}[t] x + M \frac{e^2}{2\epsilon_0 L_c^3} \left( x+\sum_{j\neq1}^{N} x_j[t]  \right)^2,
\label{potw}
\end{eqnarray}
while $G(x,q,t)+i J(x,q,t)$ is a potential that contains the remaining terms that depend on $x$, $q$, $t$, and as such carry most of the correlations,
\begin{eqnarray}
G(x,q,t)+i J(x,q,t) &=& \frac{1}{\psi^{x,q}(x,q,t)} \left[ -\frac{\hbar^2}{2m_e}\sum_{j=1,x_j\neq x_1}^{N} \left(\frac{\partial^2}{\partial x_j^2}+\frac{\partial^2}{\partial y_j^2}+\frac{\partial^2}{\partial z_j^2} \right) \psi(x,q,\mathbf{w},t)\right]_{\mathbf{w}=\mathbf{w}[t]} \nonumber \\ &+& \frac{1}{\psi^{x,q}(x,q,t)} \left[ \sum_{{ n},\lambda \neq\ 1,1}^{M,2}\frac{\hbar \omega_{n}}{2} \left(-\frac{\partial^2}{\partial q^2_{n,\lambda}} +q^2_{n,\lambda}\right) \psi(x,q,\mathbf{w},t)\right]_{\mathbf{w}=\mathbf{w}[t]} \\ &+&  \frac{i \hbar}{\psi^{x,q}(x,q,t)} \sum_{i} \left[ \frac{\partial \psi(x,q,\mathbf{w},t)}{\partial w_i} \right]_{\mathbf{w}=\mathbf{w}[t]} \frac{d w_i[t]}{dt}. \nonumber
\label{potgj}
\end{eqnarray}
If all of these potentials, together with the rest of trajectories $w_i[t]$ is known, \eqref{final0}-\eqref{final} give then the exact $\psi^{x,q}(x,q,t)$. The knowledge of the rest of trajectories can be computed similarly from other conditional wave functions. But at the practical level, however, one can only get educated guesses on the shapes of such potentials since the full many-body wavefunction is not known.

\subsubsection{Going from \eqref{final} with trajectories to \eqref{Hfull} without trajectories}

We show here the path to arrive to \eqref{Hfull} from \eqref{final}, so that one could unravel the inverse path to find ways to include some of the many-electron many-mode correlations neglected in the paper. The first simplification is to assume $G(x,q,t)+i J(x,q,t)$ in \eqref{final} as a constant, such that this constant potential profile can be ignored since it does not affect the electron velocity; this eliminates most of the many-particle correlations in our results. The second simplification is assuming $W(x,t)=V(x,t)$ as a mean-field potential that only takes into account the external bias and the RTD barrier potential, implying that we ignore both electron-electron interaction and dipole self-energy in \eqref{potw}. Notice that these terms are, in principle, easy to recover (as far as we make explicit use of trajectories) because we know the analytical shape of the potentials.

At last, the term $\sum_{j \neq 1}^{N}x_j[t]$ appearing in \eqref{final} reminds us that the conditional wave function, in principle, involves correlations with all the rest of electrons. When all $N$ electrons interact with the same single electromagnetic mode, the simplest way to take this correlation into account, while avoiding explicitly computing the trajectories, is to understand $\sum_{j \neq 1}^{N}x_j[t]$ as a sum over random positions of electrons that follow a binomial distribution, whose dispersion is given by $\sqrt{N}$, so that $(x + \sum_{j \neq 1}^{N}x_j[t]) \approx x \sqrt{N}$. Another way to understand this factor $\sqrt{N}$ is that the probabilities of the different states belonging to the optical mode are modified, not only by the electron $x \equiv x_1$  but by all the $N$ electrons; such factor $\sqrt{N}$ (instead of $N$) comes from the fact the contribution of the other electrons are not sincronized but have a random nature instead. Under all of these simplifications \eqref{final} becomes 
\begin{eqnarray}
H^{(D)}_{xq}=-\frac{\hbar^2}{2m_e}\frac{\partial^2}{\partial x^2}+ V(x,t)+\frac{\hbar \omega}{2} \left(-\frac{\partial^2}{\partial q^2} +q^2\right)+ \sqrt{2} \alpha q x ,
\label{finalfinal}
\end{eqnarray}
where the coupling constant $e \sqrt{\hbar \omega /(\epsilon_0 L_c^3)}$ for a single-electron interacting with a single-mode is then multiplied by $\sqrt{N}$, yielding $\alpha = \sqrt{e^2 \hbar \omega N /(2 \epsilon_0 L_c^3)}$, as written in \eqref{Hfull}. Such $\sqrt{N}$ is also developed elsewhere, without conditional wave function, as for example from a PZW derivation \cite{ISB_ultrastrong_THz} or from experimental grounds \cite{RTD_cavityexp}.

\end{widetext}

\section{Analytical solutions for closed two-level single-mode single-particle systems}
\label{Rabi}

The starting point in our derivation of the usual Rabi and Jaynes-Cummings models, which are basically two-level single-mode models, for a closed single-particle system is \eqref{Hfull}, but written in the energy basis of the electrons, that is, we define
\begin{eqnarray}
H_{\text{eff}} &=& \hbar \omega_0 | 0 \rangle \langle 0 |+ \hbar \omega_1 | 1 \rangle \langle 1 | + \frac{\hbar \omega}{2} (q^2 - \frac{\partial^2}{\partial q^2 }) \nonumber \\
          &+& \hbar \gamma_S \sqrt{2} q (| 0 \rangle \langle 1|  + | 1 \rangle \langle 0 |),
\label{H2l}
\end{eqnarray}
by assuming $x_{01} = \langle 0 | x | 1 \rangle = x_{10}$. The electron energies $\hbar \omega_{0,1}$ are solution of $-\hbar^2/(2 m_e) \partial^2 /\partial x^2 + V(x)$. The single-mode, level-splitting, and Rabi frequencies are respectively $\omega$, $\omega_e=\omega_1-\omega_0$, and $\gamma_S=\alpha x_{01}/\hbar$. We use a different $\gamma_S$ label here for the Rabi frequency as to differentiate from the open system label $\omega_r=\alpha L_x/\hbar$ of Section \ref{num}.

\subsection{Semiclassical Rabi model}
\label{semic}

The `standard' semiclassical approach simply considers the electromagnetic field as a time-dependent variable, with no backaction from the matter to the field. In that case, the middle term in \eqref{H2l} can be neglected as it has no effect on the purely electron dynamics, which yields the Hamiltonian of the Semiclassical Rabi model as $H_R(t) = H_R^e + H_R^\omega (t)$, where static $H_R^e$ and dynamic $ H_R^\omega (t)$ terms are
\begin{eqnarray}
H_R^e                  &=& \hbar \omega_0 | 0 \rangle \langle 0 |+ \hbar \omega_1 | 1 \rangle \langle 1 | , \label{HR0} \\
H_R^\omega (t) &=& \hbar \gamma_S \text{sin}(\omega t) (| 0 \rangle \langle 1 | + | 1 \rangle \langle 0 |),
\label{HR}
\end{eqnarray}
and where $q(t) = \text{sin}(\omega t)/\sqrt{2}$ is assumed as in the paper. The general solution $|\Psi_R(t)\rangle = c_0(t) | 0 \rangle + c_1(t) | 1 \rangle$ has the shape of \eqref{posWF} in position representation with $n=0,1$ and neglecting the photon part. 
The matrix representation of $H_R(t)$ in such a two-level basis renders $H_R^e$ as a diagonal part, while $ H_R^\omega (t)$ yields an off-diagonal part which may induce transitions - the Rabi oscillations - within the two-level subspace. The wavefunction can then be written, by factoring out the energy diagonal terms, as 
\begin{equation}
|\Psi_R(t)\rangle = c_0(t) e^{-i \omega_0 t}|0 \rangle + c_1(t) e^{-i \omega_1 t}|1 \rangle.
\label{two_WF}
\end{equation}
By using \eqref{HR0}, \eqref{HR}, \eqref{two_WF} in the Schroedinger equation $i\hbar d|\Psi_R(t)\rangle / dt = H_R(t) |\Psi_R(t)\rangle$, the coefficients evolution has analytical 
solutions \cite{scully,grynberg} given by
\begin{eqnarray}
c_0(t) &=& \bigg \{ c_0(0) \left [\text{cos}\left(\frac{\Omega_S t}{2}\right) + i\frac{\Delta}{\Omega_S}\text{sin}\left(\frac{\Omega_S t}{2}\right) \right]  \nonumber\\
          &+& c_1(0) i\frac{\gamma_S}{\Omega_S}\text{sin}\left(\frac{\Omega_S t}{2}\right) \bigg \} e^{-i\Delta t / 2}, \nonumber\\
c_1(t) &=& \bigg \{ c_1(0) \left [\text{cos}\left(\frac{\Omega_S t}{2}\right) - i\frac{\Delta}{\Omega_S}\text{sin}\left(\frac{\Omega_S t}{2}\right) \right]  \nonumber\\
	&+& c_0(0) i\frac{\gamma_S}{\Omega_S}\text{sin}\left(\frac{\Omega_S t}{2}\right) \bigg \} e^{i\Delta t / 2},
\label{C1C2}
\end{eqnarray}
where the new frequencies are the detuning $\Delta = \omega_e - \omega$ and the generalized Rabi frequency $\Omega_S = \sqrt{\gamma_S^2 + \Delta^2}$. The normalization of $|\Psi_R(t)\rangle$ implies $|c_0(t)|^2+|c_1(t)|^2=1$,
and imposes a condition on the initial values of the coefficients. So if one starts the dynamics in the excited state, $( c_0(0), c_1(0))=(0,1)$, \eqref{C1C2} simplifies to 
\begin{eqnarray}
c_0(t) &=& i\frac{\gamma_S}{\Omega_S}\text{sin}\left(\frac{\Omega_S t}{2}\right)  e^{-i\Delta t / 2}, \nonumber\\
c_1(t) &=& \left [\text{cos}\left(\frac{\Omega_S t}{2}\right) - i\frac{\Delta}{\Omega_S}\text{sin}\left(\frac{\Omega_S t}{2}\right) \right]  e^{i\Delta t / 2},
\label{C1C2_reson}
\end{eqnarray}
so that the probability of inverting the system to the ground state is $P_I(t)=|c_1(t)|^2-|c_0(t)|^2=(\Delta^2 + \gamma_S^2 \text{cos}(\Omega_S t))/\Omega_S^2$;
at resonance ($\Delta=0$) one simply has $c_0(t) = i\text{sin}(\gamma_S t / 2)$ and $c_1(t) = \text{cos}(\gamma_S t / 2)$, such that $P_I(t) = \text{cos}(\gamma_S t)$ and the system gets 
fully inverted within oscillations as given by the Rabi frequency $\gamma_S$.
Notice the crossed terms when calculating expectation values from \eqref{two_WF}; the ``energy'' for example would read
\begin{eqnarray}
\langle \Psi(t) |H_R^e + H_R^\omega(t)| \Psi(t) \rangle &=& \hbar \omega_0 |c_0(t)|^2 + \hbar \omega_1 |c_1(t)|^2 \\
                                                                     &- & 2 \hbar \gamma_S\, \mathcal{R}e\,\left[ c^{*}_1(t) c_0(t) \text{sin}(\omega t) e^{i \omega_e t}\right], \nonumber
\label{ENEt}
\end{eqnarray}
while $\langle \Psi(t) |x| \Psi(t) \rangle = 2 x_{01} \, \mathcal{R}e\left[c^{*}_1(t) c_0(t)  e^{i \omega_e t}\right]$ for the position expectation value, and 
$\langle \Psi(t) |-i\hbar d/dx| \Psi(t) \rangle = 2 p_{01}\,\mathcal{I}m\, \left[c^{*}_1(t) c_0(t)  e^{i \omega_e t}\right]$ for the momentum expectation value, where $p_{01}=\langle 0| d/dx | 1 \rangle=-\langle 1| d/dx | 0 \rangle$ is assumed.

As expected in this semiclassical picture, if no field is present ($\gamma_S=0$) and the system is initiated at the excited state as above, the system will remain in such initial state. On the other hand, the larger the coupling, the larger $\gamma_S$, and the faster the Rabi oscillations of such \textit{stimulated} emission/absorption processes. The emission (absorption) is the result of a constructive (destructive) interference among the incident radiation and the induced dipole radiation. One should mention that, in deriving \eqref{C1C2}, only one approximation is done: there appear four terms which behave like $e^{\pm i \Delta t}$ and $e^{\pm i \delta t}$, with the anti-detuning $\delta = \omega_e + \omega$. Since the latter oscillate much faster than the former, on average they yield no contribution \textit{if} one is not far from resonance and can be neglected, in which case \eqref{C1C2} is not only valid but almost exact; this is the Rotating Wave Approximation (RWA).

\subsection{Quantum Jaynes-Cummings model}
\label{quant} 

In order to derive a similar model but that keeps the quantization of the electromagnetic field, we just rewrite the field elements in \eqref{H2l} in terms of the related $\hat a$/$\hat a^\dagger$ operators, that is, we write $H_{J} = H_J^e + H_J^{\omega} + H_J^i$, with the electron (which is the same as in \eqref{HR0}, the photon, and the electron-photon interacting terms are 
\begin{eqnarray}
H_J^e                &=& \hbar \omega_0 | 0 \rangle \langle 0 |+ \hbar \omega_1 | 1 \rangle \langle 1 | , \label{HJ0} \\
H_J^{\omega}   &=& \hbar \omega  (a^\dagger a + \frac{1}{2}),  \label{HJC0} \\ 
H_J^i                &=& \hbar \gamma_S  (| 0 \rangle \langle 1| + | 1 \rangle \langle 0 |) (a + a^\dagger).
\label{HJC}
\end{eqnarray}
In \eqref{HJC0} $\hat m = a^\dagger a$ is the number of photons in the mode, which also labels the eigenstates $\{|m\rangle\}$ of $H_J^{\omega}$ with eigenenergies $\hbar \omega (m + 1/2)$. The off-diagonal terms able to induce transitions among the decoupled electron+photon states now come from the time-independent light-matter interaction term in \eqref{HJC}. Out of its four terms, $|0 \rangle \langle 1|\hat a^\dagger$ stands for a transition from excited to ground electron state via emission of a photon, while $|1 \rangle \langle 0|\hat a$ stands for the opposite process, both conserving the total electron-photon energy; on the other hand, the terms $|0 \rangle \langle 1|\hat a$ (net loss) and $|1 \rangle \langle 0|\hat a^\dagger$ (net gain) do not conserve total energy (higher order processes). \textit{If} one is not far from resonance, in the RWA spirit, these two latter terms can be neglected, so that \eqref{HJC} becomes
\begin{equation}
 H_J^i  = \hbar \gamma_S ( | 0 \rangle \langle 1 | a^\dagger+ | 1 \rangle \langle 0 | a ).
\label{HJCRWA}
\end{equation}

The general solution $|\Psi_J(t)\rangle = \sum_m (c_{0,m}(t) | 0, m \rangle + c_{1,m}(t) | 1, m \rangle)$ has the shape of \eqref{posWF} in position representation with $n=0,1$ and $m$ photons in the field. By factoring out the decoupled energy terms it reads 
\begin{eqnarray}
|\Psi_{J}(t)\rangle = &\sum_m & \big [ e^{-i(m+\frac{3}{2})\omega t} e^{-i \omega_0 t} c_{0,m+1}(t) |0,m+1 \rangle \nonumber \\
                                                       &+& e^{-i(m+\frac{1}{2})\omega t} e^{-i \omega_1 t} c_{1,m}(t)    |1,m \rangle \big ],
\label{JC_WF}
\end{eqnarray}
in which we have anticipated the fact that $H_J^i$ can only induce transitions between states $|1,m \rangle$ and $|0,m+1 \rangle$, since $\hat a^\dagger |m \rangle=\sqrt{m+1}|m+1 \rangle$ and $\hat a |m \rangle=\sqrt{m}|m-1 \rangle$. With that in mind, by using \eqref{HJ0}, \eqref{HJC0}, \eqref{HJCRWA}, \eqref{JC_WF} in the Schroedinger equation $i\hbar d|\Psi_J(t)\rangle / dt = H_J |\Psi_J(t)\rangle$, the coefficients evolution has analytical solutions \cite{scully,grynberg} given by
\begin{eqnarray}
c_{0,m+1}(t) &=& \bigg \{ c_{0,m+1}(0) \left [\text{cos}\left(\frac{\Omega_m t}{2}\right) + i\frac{\Delta}{\Omega_m}\text{sin}\left(\frac{\Omega_m t}{2}\right) \right]  \nonumber\\
   	            &-& c_{1,m}(0) i\frac{ {\gamma}_m}{\Omega_m}\text{sin}\left(\frac{\Omega_m t}{2}\right) \bigg \} e^{-i\Delta t / 2}, \nonumber\\
c_{1,m}(t)     &=& \bigg \{ c_{1,m}(0) \left [\text{cos}\left(\frac{\Omega_m t}{2}\right) - i\frac{\Delta}{\Omega_m}\text{sin}\left(\frac{\Omega_m t}{2}\right) \right]  \nonumber\\
	           &-& c_{0,m+1}(0) i\frac{ {\gamma}_m}{\Omega_m}\text{sin}\left(\frac{\Omega_m t}{2}\right) \bigg \} e^{i\Delta t / 2},
\label{JC_C1C2}
\end{eqnarray}
with the quantized Rabi frequency $ {\gamma}_m = 2\sqrt{m+1} \gamma_S$ yielding the generalized quantized Rabi frequency $\Omega_m = \sqrt{ {\gamma}_m^2 + \Delta^2}$. 
The normalization of $|\Psi_J(t)\rangle$ implies $\Sigma_m p(m) = 1$, with $p(m)=|c_{0,m}(t)|^2+|c_{1,m}(t)|^2$ the probability of $m$ photons 
in the field, while the inversion probability now becomes $P_I(t)=\Sigma_m [ |c_{1,m}(t)|^2-|c_{0,m}(t)|^2 ]$. 

Although expressions \eqref{JC_C1C2} and \eqref{C1C2} look similar, let us point out two main novelties present in the Jaynes-Cummings model as compared to the Rabi model. If one starts the dynamics in the electronic excited state, $(c_{0,m}(0), c_{1,m}(0))=(0,c_m(0))$ with $c_m(0)$ the probability amplitude of the field alone, \eqref{JC_C1C2} simplifies to 
\begin{eqnarray}
c_{0,m+1}(t) &=& -c_{m}(0) i\frac{ {\gamma}_m}{\Omega_m}\text{sin}\left(\frac{\Omega_m t}{2}\right) e^{-i\Delta t / 2}, \label{JC_C1C2_reson} \\
c_{1,m}(t)     &=&   c_{m}(0) \left [\text{cos}\left(\frac{\Omega_m t}{2}\right) - i\frac{\Delta}{\Omega_m}\text{sin}\left(\frac{\Omega_m t}{2}\right) \right] e^{i\Delta t / 2}, \nonumber
\end{eqnarray}
so that $P_I(t) = \Sigma_m |c_m(0)|^2 [(\Delta^2 + \gamma_m^2 \text{cos}(\Omega_m t))/\Omega_m^2]$. One can envision
two cases: i) for an initial vacuum field, that is, $|c_m(0)|^2=\delta_{m,0}$, and under resonance ($\Delta = 0$), one simply has the coefficients 
$c_{0,1}(t) = - i \text{sin}( \gamma_0 t/2)$ and $c_{1,0}(t) = \text{cos}( \gamma_0 t/2)$, so that $P_I(t) = \text{cos}(\gamma_0 t)$ and
the system can be fully inverted, even without an external field, due to \textit{spontaneous} photon emission; ii) by using a typical coherent initial Poisson distribution 
around the average photon number $\langle m \rangle$, that is, $|c_m(0)|^2=\langle m \rangle ^m e^{-\langle m \rangle}/m!$, as $\langle m \rangle$
increases, one obtains collapses and revivals of $P_I(t)$ as caused by interferences among the different terms in the sum over $m$, which oscillate at their
own distinct frequencies. Notice that, although a continuous photon distribution or even a classical random field could also induce a collapse in $P_I(t)$, the revivals 
are a pure quantum feature. We are here interested in case i).

As next subsection will let clear, the light-matter coupling creates a doublet structure in the spectrum of the full Hamiltonian $H_J$. If one focus on an initial vacuum field, that is, $m=0$ in \eqref{JC_WF}, one sees that the lowest doublet is given by
\begin{equation}
|\Psi_J(t)\rangle =   e^{-i\frac{3\omega t}{2}} e^{-i \omega_0 t} c_{0,1}(t) |0,1 \rangle + e^{-i\frac{\omega t}{2}} e^{-i \omega_1 t} c_{1,0}(t) |1,0 \rangle,
\label{JC_WF_doublet}
\end{equation}
which couples the excited eletron state without photon $|1,0 \rangle$ to the ground electron state with a single photon $|0,1 \rangle$, while $|0,0 \rangle$ is unnafected and $|1,1 \rangle$ would only couple to $|0,2 \rangle$ in the next doublet. When the system is initiated in the $|1,0 \rangle$ state one simply has, from \eqref{JC_C1C2},
\begin{eqnarray}
c_{0,1}(t) &=&  i\frac{ \gamma_0}{\Omega_0}\text{sin}\left(\frac{\Omega_0 t}{2}\right) e^{-i\Delta t / 2}, \label{JC_C1C2_01} \\
c_{1,0}(t) &=&  \left [\text{cos}\left(\frac{\Omega_0 t}{2}\right) - i\frac{\Delta}{\Omega_0}\text{sin}\left(\frac{\Omega_0 t}{2}\right) \right] e^{i\Delta t / 2}. \nonumber
\end{eqnarray}

Notice that \eqref{JC_WF_doublet} and \eqref{JC_C1C2_01} resemble \eqref{two_WF} and \eqref{C1C2_reson}. Expectation values however have distinct features with regard to the semiclassical case. The energy becomes
\begin{eqnarray}
&&\langle \Psi(t) |H_J^e + H_J^{\omega} + H_J^i| \Psi(t) \rangle = (\hbar \omega_0 + 3\hbar \omega/2)|c_{0,1}(t)|^2  \label{JC_ENEt} \\
&&  + (\hbar \omega_1 + \hbar \omega/2) |c_{1,0}(t)|^2 -  2 \hbar \gamma_0 \; \mathcal{R}e \left[ c^{*}_{0,1}(t) c_{1,0}(t) e^{-i \Delta t}\right], \nonumber
\end{eqnarray}
which, at resonance, is conserved as expected since the crossed term vanishes; at $\Delta > 0$ tough the same may not be true. Both position and momentum expectation values for the electron vanish, $\langle \Psi(t) |x| \Psi(t) \rangle=\langle \Psi(t) |-i\hbar d/dx| \Psi(t) \rangle=0$, due to the orthogonality of the photon states. And the same happens for their photon counterparts, $\langle \Psi(t) |q| \Psi(t) \rangle=\langle \Psi(t) |-i\hbar d/dq| \Psi(t) \rangle=0$, due to the orthogonality of the electron states. Only bilinear terms involving $xq$ (or $d/dx \; \; d/dq$), at $\Delta > 0$, may be non-zero, since $\langle \Psi(t) |xq| \Psi(t) \rangle=2 x_{01} q_{01} \; \mathcal{R}e \left[ c^{*}_{0,1}(t) c_{1,0}(t) e^{-i \Delta t} \right]$, assuming $q_{01}=\langle 0| q | 1 \rangle=q_{10}$. As expected, these last terms in the Real part are the same ones found in \eqref{JC_ENEt}, since the interaction Hamiltonian in \eqref{HJCRWA} is derived from such a bilinear term.

\subsection{Light-dressed states of the matter: Polaritons}
\label{dressed}

Having in mind that in the Jaynes-Cummings model the full Hamiltonian $H_J$ is time independent, and that under RWA approximation $H_J^i$ can only induce $|0,m+1 \rangle\leftrightarrow|1,m \rangle$ transitions in the decoupled $H_J^e+H_J^{\omega}$, such that state $|0,0 \rangle$ is unaffected, one realizes that the spectrum of $H_J$ is given by a set of energy doublets $D_m$ \textit{if} one is not far from resonance; that is, $D_1$ contains states $|0,1 \rangle$-$|1,0 \rangle$, $D_2$ builds on $|0,2 \rangle$-$|1,1 \rangle$, and so on. As such, the Hamiltonian $H_{D_m}$ of each doublet subspace
\begin{equation}
H_{D_m} = \hbar
\begin{bmatrix}
\omega_0 + (m+3/2)\omega & -\gamma_m \\
-\gamma_m               & \omega_1 +  (m+1/2)\omega  \\
\end{bmatrix}
\label{Hmat}
\end{equation}
can easily be diagonalized, yielding the eigenvalues $E_{\pm,m}$, 
\begin{equation}
E_{\pm,m}= \hbar((\omega_0+\omega_1)/2 + (m+1)\omega \pm \sqrt{(\Delta/2)^2+\gamma_m^2}), 
\label{eigenH_ener}	
\end{equation}
and the eigenvectors $|\psi_{\pm,m}\rangle$,
\begin{eqnarray}
|\psi_{-,m} \rangle  &= &-\text{sin}(\theta_m/2)|0,m+1 \rangle + \text{cos}(\theta_m/2)|1,m \rangle, \nonumber \\
|\psi_{+,m} \rangle &=& \text{cos}(\theta_m/2)|0,m+1 \rangle + \text{sin}(\theta_m/2)|1,m \rangle,
\label{eigenH}
\end{eqnarray}
with $\text{tan} (\theta_m) = 2|\gamma_m|/\Delta$ ($0 \le \theta_m < \pi$). Notice that $E_{+,m} - E_{-,m} = 2\hbar \sqrt{(\Delta/2)^2+\gamma_m^2}$, so that even at resonance ($\Delta=0$, $\theta_m = \pi/2$), where the decoupled levels of $H_J^e+H_J^{\omega}$ are degenerate, the coupling term $H_J^i$ induces an avoided crossing in each doublet $D_m$, and creates fully entangled light-matter states,
\begin{equation}
|\psi_{\pm,m}^{(\Delta=0)}\rangle = \frac{1}{\sqrt{2}}[ |1,m \rangle \pm |0,m+1 \rangle ],
\end{equation} 
which are the so-called light-dressed states of the matter, or polaritons. As $\Delta$ increases so that $\theta_m$ approaches $0$ or $\pi$, $|\psi_{\pm,m} \rangle$ tends to the uncorrelated bare states $|0,m+1 \rangle$ or $|1,m \rangle$. So the Rabi oscillations can be understood from the polaritonic picture: one inverts \eqref{eigenH} to have $|0,m+1 \rangle$,$|1,m \rangle$ in terms of $|\psi_{\pm,m}\rangle$ and, if at $t=0$ one starts the dynamics around $\Delta=0$ at $|1,m \rangle$, which is not an stationary state $|\psi_{\pm,m} \rangle$ of the full $H_J$, at a later time the dynamics will bring the system to $|0,m+1 \rangle$.

In the case one deals with an isolated system defined by some confining potential (Appendix \ref{appendixres} considers an infinite well), with known electron eigenstates $\phi_n^{(e)}(x)$ in \eqref{posWF}-\eqref{posWF2_e}, one can project 
\eqref{posHproj} in $\phi_j^{(e)}(x)$ as to also integrate out the electron degree of freedom, resulting
\begin{eqnarray}
i\hbar \frac{d}{d t} c_{j,k}(t) &=& \left[ E_j + \hbar \omega (k + 1/2) \right] c_{j,k}(t) \label{poscjk} \\
          	                                     &+& \alpha \sum_n \left[ \sqrt{k+1} c_{n,k+1} + \sqrt{k} c_{n,k-1} \right] \langle j |x| n \rangle. \nonumber
\end{eqnarray}
Usual confining potentials will have symmetries like parity dictating selection rules for real dipole matrix elements, so that $ \langle j |x| n \rangle = \langle n |x| j \rangle = x_{j,n} (\delta_{j,n+1} + \delta_{j,n-1})$, that is, only two neighboring electronic states are connected, and as such \eqref{poscjk} yields
\begin{eqnarray}
i\hbar \frac{d}{d t} c_{j,k}(t) &=& \left[ E_j + \hbar \omega (k + 1/2) \right] c_{j,k}(t) \label{poscjk_sym} \\
	                                     &+& \alpha \sqrt{k+1} \left[x_{j,j-1} c_{j-1,k+1} + x_{j,j+1} c_{j+1,k+1}\right] \nonumber \\
	                                     &+& \alpha \sqrt{k} \left[x_{j,j-1} c_{j-1,k-1} + x_{j,j+1} c_{j+1,k-1}\right]. \nonumber
\end{eqnarray}
By employing the RWA and so neglecting the terms  $c_{j-1,k-1}$ and  $c_{j+1,k+1}$ that do not conserve the number of electron-photon excitations, \eqref{poscjk_sym} becomes
\begin{eqnarray}
i\hbar \frac{d}{d t} c_{j,k}(t) &=& \left[ E_j + \hbar \omega (k + 1/2) \right] c_{j,k}(t) \label{poscjk_RWA} \\
	                                     &+& \alpha \sqrt{k+1} x_{j,j-1} c_{j-1,k+1}  + \alpha \sqrt{k} x_{j,j+1} c_{j+1,k-1}. \nonumber
\end{eqnarray}
As an example, in the case of a two-level system for the electrons ($j=0,1$), for the two lowest photon states ($k=0,1$), \eqref{poscjk_RWA} yields
\begin{eqnarray}
i\hbar \frac{d}{d t} c_{0,0}(t) &=& \left[ E_0 + \hbar \omega/2 \right] c_{0,0}(t), \label{poscjk_RWA2} \\
i\hbar \frac{d}{d t} c_{1,0}(t) &=& \left[ E_1 + \hbar \omega/2 \right] c_{1,0}(t) + \alpha x_{0,1} c_{0,1}(t), \nonumber \\
i\hbar \frac{d}{d t} c_{0,1}(t) &=& \left[ E_0 + 3\hbar \omega/2)\right] c_{0,1}(t) + \alpha x_{0,1} c_{1,0}(t), \nonumber \\
i\hbar \frac{d}{d t} c_{1,1}(t) &=& \left[ E_1 + 3\hbar \omega/2) \right] c_{1,1}(t) + \alpha \sqrt{2} x_{0,1} c_{0,2}(t), \nonumber
\end{eqnarray}
where in the last equation the term $\alpha x_{1,2} c_{2,0}(t)$ was not included since it refers to the next electron state $j=2$. So, as expected in an isolated system, we recover the results from the Jaynes-Cummings model since the equations for $c_{1,0}(t)$ and $c_{0,1}(t)$ are the same obtained in the development of \eqref{JC_C1C2}, having those same solutions. They constitute the doublet $D_1$ in \eqref{JC_C1C2_01}, and as we have just seen $c_{0,0}(t)$ is decoupled, while  $c_{1,1}(t)$, by coupling with  $c_{0,2}(t)$, would build the next doublet $D_2$. We emphasize, however, that \eqref{posHproj} is more general for dealing with open transport scenarios, since it does not dwell on the specific information of an isolated system basis.

\section{Results for Rabi, Jaynes-Cummings, and Bohmian scenarios in isolated systems}
\label{appendixres}

For some sort of comparison among these closed analytical models and the open transport scenario from the main text, we consider here a $1$D infinite symmetric quantum well potential with similar parameters from the active region of our RTD, that is, $m_e=0.041$ $m_0$ and $L_x = 16$ nm. The ground and first excited electronic basis functions are $\phi_0^{(e)}(x) = \sqrt{2/L_x}\text{cos}(\pi x/L_x)$ and $\phi_1^{(e)}(x) = -\sqrt{2/L_x}\text{sin}(2\pi x/L_x)$, yielding $x_{01} \approx -2.9$ nm and $p_{01} \approx -0.11$ nm$^{-1}$. The respective eigenenergies are $\hbar \omega_0 \approx 35$ meV and $\hbar \omega_1 \approx 140$ meV, yielding $\omega_e \approx 160$ Trad/s and, for a resonant field $\omega=\omega_e$, $\nu=\omega / (2 \pi)\approx 25$ THz and $\hbar \omega/2\approx 52$ meV. For the quantum case in the vacuum field, the zero-photon and single-photon basis functions are $\phi_0^{(\omega)}(q)=(1/(\pi l_0^2))^{1/4}e^{-q^2/(2 l_0^2)}$ and $\phi_1^{(\omega)}(q)=(1/(\pi l_0^2))^{1/4}e^{-q^2/(2 l_0^2)} \sqrt{2} q/l_0$, with $l_0=\sqrt{\hbar/\omega}$. When $\Delta = 0$, $\Omega_S = \gamma_S$ for the semiclassical scenario in \eqref{two_WF}, \eqref{C1C2_reson}, while $\Omega_0 = \gamma_0$ for the quantum scenario in \eqref{JC_WF_doublet}, \eqref{JC_C1C2_01}. The dynamics starts from the excited electron state (without photon in the quantum case). In practical terms, the Rabi frequency can be taken as a parameter 
- smaller than $\omega$ for the validity of the models -, dictating the intensity of the light-matter coupling; for a meaningful comparison we then define $\tilde \gamma = \gamma_0 = 2 \gamma_S$, and take $\tilde \gamma = \omega / 5 \approx 32$ Trad/s (the same value employed for $\omega_r$ in Section \ref{num}), which here yields $\zeta \equiv \tilde \gamma / \omega \approx 1/5$. This is already a large value of $\zeta$ close to the limits of validity of the present analytical models.

\begin{figure}
\includegraphics[scale=0.1]{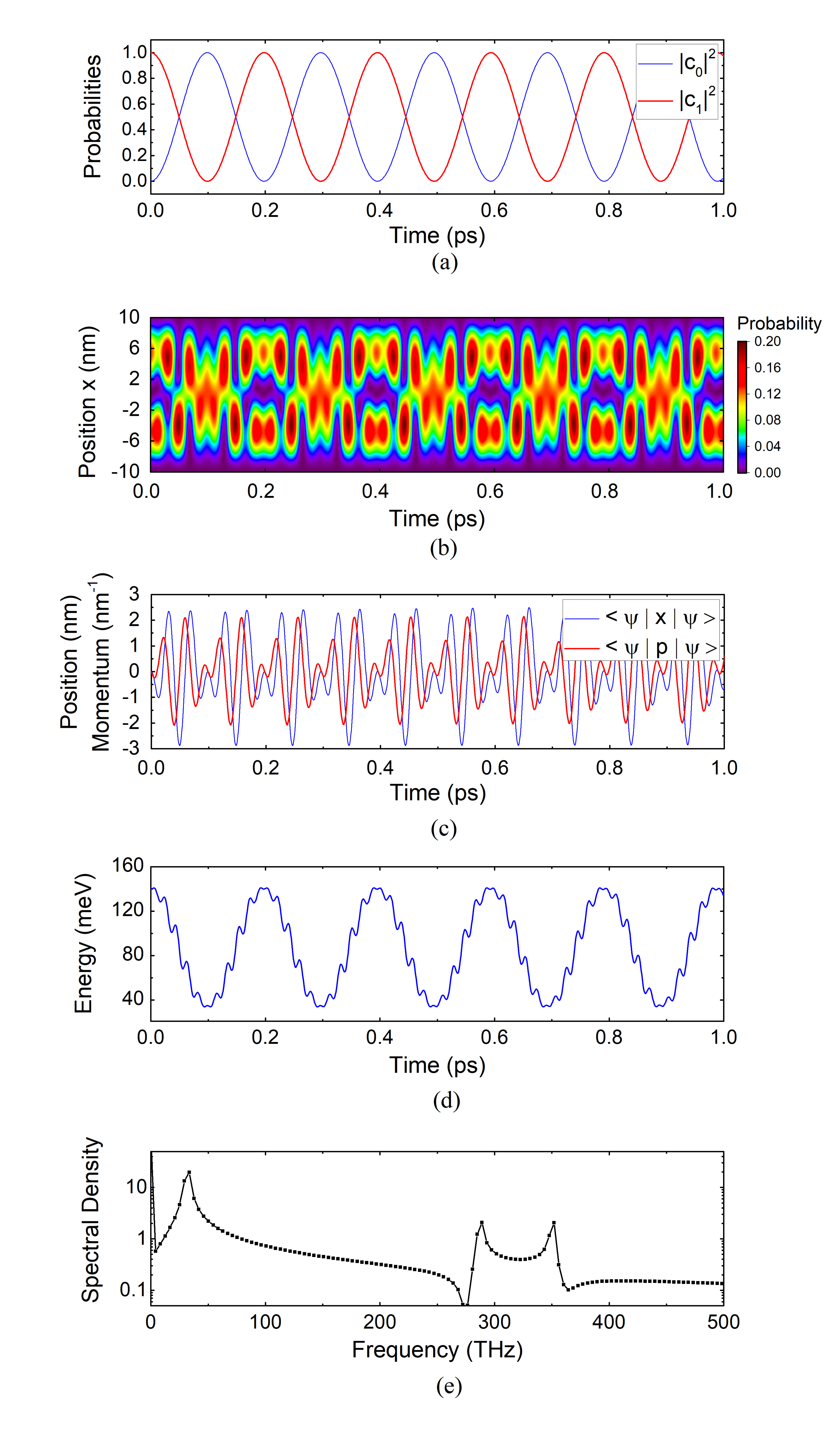}
\caption{Time-evolution of the semiclassical Rabi model. (a) ground and excited state probabilities; (b) wavefunction distribution $|\Psi_R(t)|^2$; (c) expectation values of position $\langle \Psi_R(t) |x| \Psi_R(t) \rangle$ and momentum $\langle \Psi_R(t) |p| \Psi_R(t) \rangle$ (amplified by $20$); (d)``energy'' $E(t)=\langle \Psi_R(t) |H_R^e+H_R^{\omega}(t)| \Psi_R(t) \rangle$; (e) spectral density.}
\label{SC}
\end{figure}

\begin{figure}
\includegraphics[scale=0.1]{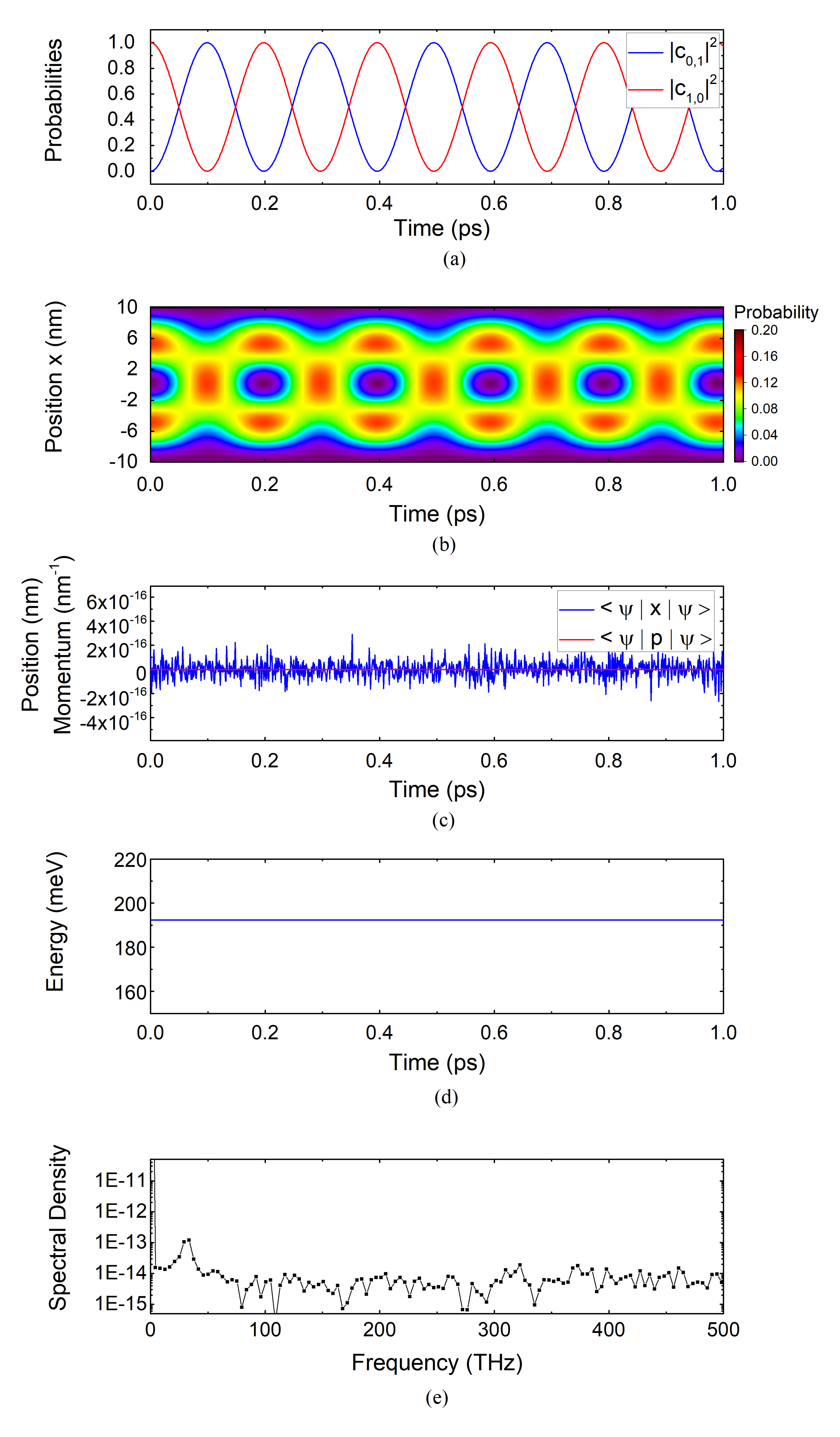}
\caption{Time-evolution of the quantum Jaynes-Cummings model. (a) ground and excited state probabilities; (b) wavefunction distribution $|\Psi_J(t)|^2$; (c) expectation values of position $\langle \Psi_J(t) |x| \Psi_J(t) \rangle$ and momentum $\langle \Psi_J(t) |p| \Psi_J(t) \rangle$; (d) energy $E=\langle \Psi_J(t) |H_J^e+H_J^{\omega}+H_J^i| \Psi_J(t) \rangle$; (e) spectral density.}
\label{JC}
\end{figure}

Figures \ref{SC} and \ref{JC}, each with five panels, deal respectively with the semiclassical Rabi model and with the quantum Jaynes-Cummings model, both in resonant situation. Results in general show what one expects from the discussions so far, with Rabi oscillations within period of $2\pi/ \tilde \gamma \approx 0.2$ ps. Panels (a) in both figures present a periodic full inversion of the system, from the respective excited to ground states. Panels (b) show a distinct look among semiclassical $|\Psi_R(x,t)|^2$ and quantum $|\Psi_J(x,t)|^2$; this happens because, while the phases $e^{i\omega_{0,1}t}$ in the $1$D equation in \eqref{two_WF} create the `noisy' semiclassical pattern due to the crossed terms, in the quantum pattern the photon degree of freedom is integrated out in the $2$D equation in \eqref{JC_WF_doublet} so that its orthogonality eliminates crossing terms. Notice tough that, had we considered the semiclassical case without those phases its evolution would look exactly the same as the quantum case, alternating between double (excited state) to single (ground state) peaks.
 
\begin{figure}
\includegraphics[width=0.9\linewidth]{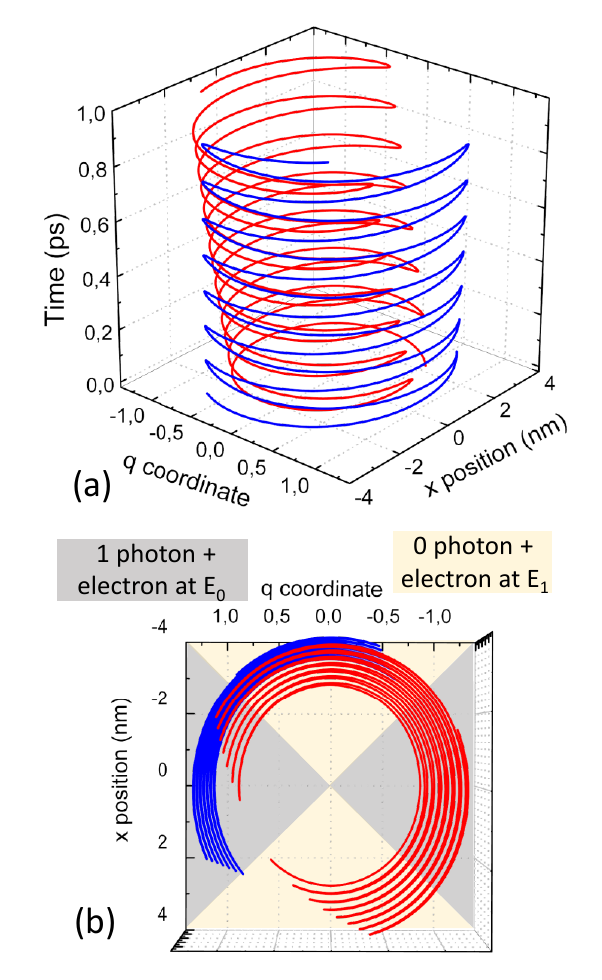}
\caption{(a) Example of two Bohmian trajectories in the $2$D $xq$ plane guided by the analytical evolution from the quantum scenario. (b) Schematic representation of different regions on the $xq$ configuration space in the quantum well, with white (grey) regions corresponding to the wavefunction (where the values $x[t]$ and $q[t]$ have higher probability) occupying the zero (one) photon and excited (ground) electron energy state.}
\label{Traj_analytical}
\end{figure}

In panel (c), for the semiclassical model, the expectation value for the position $\langle \Psi_R(t) |x| \Psi_R(t) \rangle$ oscillates in between $\pm x_{01}$, while for the momentum $\langle \Psi_R(t) |-i\hbar d/dx| \Psi_R(t) \rangle$ oscillates in between $\pm \hbar p_{01}$ (scaled by $20$ in the figure); one sees that frequencies other than the Rabi one are playing a role in the dynamics. On the other hand, in panel (c) for the quantum model, due to the orthogonality of the photon degree, both expectation values $\langle \Psi_J(t) |x| \Psi_J(t) \rangle$ and $\langle \Psi_J(t) |-i\hbar d/dx| \Psi_J(t) \rangle$ are zero; similarly, had we plotted $\langle \Psi_J(t) |q| \Psi_J(t) \rangle$ and $\langle \Psi_J(t) |-i\hbar d/dq| \Psi_J(t) \rangle$ the same would happen due to the orthogonality of the electron degree. The energy expectation value in panel (d) for the quantum picture, $E=\langle \Psi_J(t) |H_J^e+H_J^{\omega}+H_J^i| \Psi_J(t) \rangle$, which comes from a time-independent Hamiltonian, remains conserved at the initial excited electron plus zero photon energy ($\approx 140 + 52 \approx 192$ meV). On the other hand, the ``energy'' expectation value for the semiclassical picture, $E=\langle \Psi_R(t) |H_R^e+H_R^{\omega}(t)| \Psi_R(t) \rangle$, follows the expected trend with the Rabi oscillations, remaining in between excited ($\approx 140$ meV) and ground ($\approx 35$ meV) electron state energies, also with a clear signature of additional frequencies. From a Fourier transform of the energies, as shown in panels (e), the quantum case obviously has no peaks, while in the semiclassical case one clearly identifies the first peak at the Rabi frequency $\tilde \gamma \approx 32$ Trad/s, while the other two peaks are located at $2\omega - \tilde \gamma \approx 288$ Trad/s and $2\omega + \tilde \gamma \approx 352$ Trad/s, being separated by $2 \tilde \gamma$ as explained in previous Appendix.

When comparing the numerical results from our RTD open system with the analytical results from an `analogous' closed system, one notices that the semiclassical scenario in Figs. \ref{Numerical_quant_probs_alpha25_lev2}(b) and \ref{SC}(b) is always noisier with regard to the quantum scenario in Figs. \ref{Numerical_quant_probs_alpha25_lev2}(c) and \ref{JC}(b) due to extra frequencies. Obviously both closed systems present a lasting coherent evolution dictated by the Rabi frequency, while in both open systems the probabilities decrease upon time inside the active region due to transmission and reflection of the wavepackets. 

To exemplify the use of the Bohmian trajectories as developed in \eqref{ap_traj}, we show in Fig. \ref{Traj_analytical} an example of two of such trajectories in the $2$D $xq$ space for the quantum picture from Fig. \ref{JC}(b); trajectories in the $1$D $x$ space for the semiclassical picture are straightforward and not shown. Their time evolution follows a spiraling path as seen in Fig. \ref{Traj_analytical}(a), which is the signature of the Rabi oscillation: i) electron in the first excited state and zero photon in the cavity, trajectory in the white area of Fig. \ref{Traj_analytical}(b); ii) electron in the ground state and one photon in the cavity, trajectory in the grey area of Fig. \ref{Traj_analytical}(b). The turning points of each oscillation, when the trajectory reaches null velocity, always alternate from one quadrant to the other, in different ways depending on the initial conditions of each trajectory. This transition between different quadrants is a signature of the change of the dressed system between the two eigenstates, and underlines how this evolution is described in a continuous way.

While the values of $q$ remain in between $\pm 1$, the values of $x$ remain in between $\pm x_{01}$. If trajectories for the open system were plotted, oscillations would be seen, at small $t$, for values of $x$ in between $\pm L_x$; at large $t$, trajectories would be going `chaotically'  away from the active region due to transmission or reflection of the wavepackets.

\end{appendix}

\end{document}